\documentclass[11pt]{article}

\usepackage{epsf,epsfig,amsmath,amsfonts}
\usepackage{graphicx}

\usepackage{a4wide}
\usepackage{amsmath}
\usepackage{amscd}
\usepackage{amssymb}
\usepackage{amsthm}
\usepackage{amsmath}
\usepackage{latexsym}
\usepackage{bbm}
\usepackage{cases}
\usepackage{graphicx}
\usepackage{subfig}
\usepackage{hyperref}
\usepackage{natbib}
\usepackage{curves}
\usepackage{xcolor}
\usepackage[left=40mm, right=20mm]{geometry}

\usepackage{setspace}
\newtheorem{thm}{Theorem}[section]
\theoremstyle{plain}
\newtheorem{lem}[thm]{Lemma}
\newtheorem{prop}[thm]{Proposition}

\theoremstyle{definition}
\newtheorem{defi}[thm]{Definition}
\newtheorem{rem}[thm]{Remark}
\newtheorem{assum}[thm]{Assumption}

\newtheorem{ex}{Example}

\newcommand{\ktP}{\ensuremath{\mathbb{P}}}
\newcommand{\ktR}{\ensuremath{\mathbb{R}}}

\newcommand{\htG}{\ensuremath{\mathcal{G}}}

\newcommand{\gd}{c\`{a}gl\`{a}d }
\newcommand{\dg}{c\`{a}dl\`{a}g }

\numberwithin{equation}{section}
\title{Optimal liquidation for a risk averse investor in a one-sided limit order book driven by
a L\'{e}vy process}

\date{\today}

\author{Arne L\o kka\footnote{
    Department of Mathematics
    Columbia House
    London School of Economics
    Houghton Street, London WC2A 2AE
    United Kingdom
    (a.lokka\@@lse.ac.uk)}
    \and
    Junwei Xu\footnote{
    Department of Mathematics
    Columbia House
    London School of Economics
    Houghton Street, London WC2A 2AE
    United Kingdom
    (j.xu19\@@lse.ac.uk)}}
\begin{document}
\maketitle

\begin{abstract}
In a one-sided limit order book, satisfying some realistic assumptions, where
the unaffected price process follows a L\'{e}vy process, we consider a market agent
that wants to liquidate a large position of shares. We assume that the agent has
constant absolute risk aversion and aims at maximising the expected utility of the
cash position at the end of time. 
The agent is then faced with the problem of
balancing the market risk and the cost of a rapid execution. In particular we are interested in
how the agent should go about optimally submitting orders.
Since liquidation normally takes place within a short period of time, modelling the risk as a L\'{e}vy process
should provide a realistic model with good statistical fit to observed market data, and thus the model 
should provide a realistic reflection of the agent's market risk.
%Our formulation also allows for a discontinuous limit order book which can provide a reasonable approximation 
%for a limit order book with a discrete shape in practice. 
We reduce the optimisation problem to a deterministic two-dimensional singular problem,
to which we are able to derive an explicit solution in terms of the model data. In particular we find an expression for the 
optimal intervention boundary, which completely characterise the optimal liquidation strategy.
%In particular, this problem provides an example of a solvable two-dimensional singular Markovian optimal 
%control problem with an optimal intervention boundary which could be discontinuous.
\end{abstract}

\section{Introduction}

This paper is concerned with how a market agent should go about selling
(or purchasing) a large position of shares. This kind of problem has attracted
considerable interest over the past few years following the introduction of
electronic trading platforms. In the model we consider, we specify the limit
order book and how this recovers over time. Thus the optimal liquidation
strategy will explicitly specify the orders the agent submits
to the market, as opposed to just specifying the optimal speed at which to
trade that is the case for the popular impact models. We refer the reader
to \citet{LL}, 
\citet{aCJP} and \citet{Gu} for an introduction to optimal execution and common 
models.

%Optimal liquidation studies the problem that a large investor aims to sell 
%a certain huge amount of asserts. Due to 
%the large trading volume from this investor, his strategy affects market price. 
%This can potentially result in significant execution costs. Therefore he 
%tries to find out an optimal way of selling. How to model the influence 
%on market price caused by the large investor's trading is a crucial aspect 
%of dealing with an optimal liquidation problem. 

More precisely, we consider a market agent with constant absolute risk aversion (CARA)
that wants to maximise the expected utility of the cash position at the end of time.
Thus the agent does not face any restrictions on the duration of the liquidation, but the
rapidness is determined by the market risk and the agent's preference towards risk.
Working with an infinite time-horizon also avoids the time dimension of the problem, and
hence make the problem more tractable. 
We assume that the market risk of the stock price is modelled by
a L\'{e}vy process, which is allowed to have a drift, but which we assume satisfy a certain
exponential moment condition. A number of studies demonstrate that L\'{e}vy 
processes are able to capture the essential statistical properties of stock price movements
over short time-horizons \citep{MS}, \citep{EK}, \citep{Barn} and \citep{CT}.
Since the main bulk of the liquidation tends to finish within a short period of time, this model should provide
a reasonable reflection of the market risk faced by the agent.
For the same reason, a L\'{e}vy process model is a reasonable approximation of 
an exponential L\'{e}vy process model. 
Such a Bachelier-L\'{e}vy type model may seem simplistic, 
but this kind of modelling of the unaffected price process is widely used 
in the optimal liquidation literature \citep{AC2}, \citep{KM}, \citep{SS} and \citep{Gath}. 
%There are studies which show that liquidation models with linear price processes 
%provide a good approximation to models with exponential price processes 
%\citep{GS} and \citet{FKTW}. 
In particular, \citet{FKTW} demonstrated that the linear model provides an excellent approximation to 
models with prices modelled as a geometric Brownian motion and multiplicative impact in the 
Almgren-Chriss framework.

We consider a bid limit order book with general shape and with a general resilience function
satisfying some reasonable conditions. In particular, some of these assumptions are crucial
in order to solve the problem since the assumptions imply that a certain function is concave,
which is needed for optimality of our strategy. 
%NOTE
%It is assumed that the 
%unaffected bid price provides a lower bound for the best ask price 
%and that the bid limit order book is unaffected by the large investor's buy orders. 
%These assumptions allow us to exclude any buy orders in the optimal trading strategy, 
%and they also exclude price manipulation in our model in the sense of \citet{HS}. NOTE
%The available number of bid limit orders are assumed to be finite, which limits the 
%agent's trading strategy in the sense that the agent can not make a block sale larger
%than currently available number of bid orders. 
So with an infinite time horizon, we solve the problem of maximising the expected utility of the 
agent's final cash position. We do this by showing that the problem can be reduced
to a two-dimensional deterministic singular optimisation problem to which we can obtain an explicit solution
in terms of the characteristics of the limit order book and the agent's risk aversion.
With reference to \citet{Lokk} and the nature of the problem, we guess
that the optimal strategy consists initially of either a block sale or a period of 
waiting, and that the  agent thereafter continuously submit sales orders in such a way that 
the state process remains on the  optimal intervention boundary. The state process 
here is the two-dimensional process consisting of the number of shares
the agent currently hold and the current state of the order book.  
The optimal intervention boundary is associated with the Hamilton-Jacobi-Bellman (HJB) 
variational inequalities corresponding to the optimisation problem. 
This intervention boundary might have discontinuities as well as constant parts.
The discontinuities corresponds to periods of waiting while the order book recovers, while the constant
part corresponds to submitting sales orders at the same rate as the resilience rate.
%The appearance of jumps 
%of intervention boundary is a comprehensive consequence due to the cumulant 
%generating function of the L\'{e}vy process, shape of limit order book and mode of 
%resilience rate.
%Moreover, the intervention boundary is non-increasing. This means that 
%when the agent is continuously selling shares, it is never optimal to implement a 
%speed which is greater than the resilience rate. 
Following the idea in \citet{Lokk}, 
the value function in our problem can be expressed in an explicit way in terms of the
problem's data, and we 
characterise the intervention boundary via the HJB variational inequalities. 
The strategy associated with this intervention boundary is shown to 
be optimal by a verification argument. 

We also provide an example in order to illustrate the optimal strategies for various parameters
of risk aversion for the case when the risk is modelled by a Brownian motion and for the
case when the risk is modelled by a L\'{e}vy process with jumps. For the case of a L\'{e}vy process
with jumps we choose the L\'{e}vy process approximation of the exponential variance-gamma
process. We find that the two models produce similar optimal strategies if the agent is not very
risk averse, but as the agent's stock position and the agent's risk aversion increase, the
differences become more pronounced.

The model we use is a version of the model introduced in \citet{OW}, which was later generalised
in \citet{AFS}, and then further in \citet{PSS}.
However, these papers did not consider risk-aversion, but assume that the agent wants to 
maximise the expected value of the cash position.
The problem we consider in this paper is an extension of \citet{Lokk} in the sense that unaffected price 
process follows a general L\'{e}vy process which could have a drift, and not simply a Brownian motion.
Moreover, in this paper we consider an order book with a general resilience function, not just exponential
resilience.

This paper is structured as follows. In Section 2 we introduce the limit 
order book model and the agent's optimisation problem. We simplify the problem and 
show that it can be reduced to a deterministic optimisation problem in Section 3.
The simplified deterministic optimisation problem is solved in Section 4. The proofs omitted in the previous sections
are contained in Section 5.

\section{Problem formulation}

Let $\bigl(\Omega, \mathcal{F}, (\mathcal{F}_t)_{t\geq 0},\mathbb{P}\bigr)$ 
be a complete filtered probability space satisfying the usual conditions and 
supporting a one-dimensional non-trivial L\'{e}vy process $L$. 
%We adopt the convention that $\mathcal{F}_{0-}=\mathcal{F}_0$ and $L_{0-}=L_0=0$. 
\begin{assum}\label{AssumpL_LOB}
We assume that there exists some $\delta>0$ such that $\mathbb{E}\bigl[\mathrm{e}^{\theta L_1}\bigr]<\infty$, 
for $\arrowvert\theta\arrowvert<\delta$.
\end{assum}

Let $\kappa$ denote the cumulant generating function of $L_1$, i.e.
\begin{gather*}
\kappa(\theta)=\ln\bigl(\mathbb{E}\bigl[\mathrm{e}^{\theta L_1}\bigr]\bigr) ,
\qquad\theta\in\mathbb{R} .
\end{gather*}
Assumption \ref{AssumpL_LOB} guarantees that the cumulant generating 
function $\kappa$ is continuously differentiable in a neighbourhood 
of $0$ and that the L\'{e}vy process $L$ is square integrable. Hence $L$ admits the representation
\begin{gather*}
L_t=\mu t+\sigma W_t+\int_{\mathbb{R}\setminus\{0\}}
z \bigl(N(t,dz)-t\nu(dz)\bigr), \qquad t\geq 0,
\end{gather*}
where $\mu\in\mathbb{R}$, $W$ is a standard Brownian motion, $N$ is a Poisson random 
measure which is independent of $W$ with compensator $\pi(t,dz)=t\nu(dz)$, where
$\nu$ denotes the L\'{e}vy measure associated with $L$ \citep{Kypr}. 
The 
cumulant generating function $\kappa$ can then be expressed as
\begin{gather} \label{cumulant_func}
\kappa(\theta)=\mu\theta+\frac{1}{2}\sigma^2\theta^2
+\int_{\mathbb{R}\setminus\{0\}}
\Bigl(\mathrm{e}^{\theta z}-1-\theta z\Bigr)\, \nu(dz) ,
\qquad \arrowvert \theta\arrowvert<\delta .
\end{gather}
In particular,
\begin{gather*}
\kappa(0)=0,\qquad\kappa'(0)=\mu \qquad\text{ and }\qquad
\kappa''(0)=\sigma^2+\int_{\mathbb{R}\setminus\{0\}} z^2\nu(dz). %\label{kappa''}
\end{gather*}
Moreover, $\kappa$ is 
%lower semi-continuous \citep[see][Lemma 2.3]{GOW} and 
strictly convex and continuously differentiable on its effective domain. 

\begin{assum}\label{supermart}
We assume that $\mu\leq 0$, i.e.\,the L\'{e}vy process $L$ is a supermartingale.
\end{assum}
The main reason behind this assumption is that the optimisation problem we consider in this paper does not have a solution when
$\mu$ is positive. This is because when $\mu$ is positive, it is be optimal for the agent to hold on to some amount of shares 
for as long as possible, and as a consequence our formulation of the liquidation problem does not have a solution when $\mu$
is positive (see Remark \ref{muposrem}). 
For the remainder of the paper, we therefore restrict ourselves to the case $\mu\leq 0$ for which our optimisation problem is well formulated.
To simplify notation set
\begin{gather*}
\mathbb{R}^+=[0,\infty)\quad\text{ and }\quad
\mathbb{R}^-=(-\infty,0] .
\end{gather*}
%Therefore, $\kappa(\theta)$ 
%is strictly decreasing for $\theta<0$ and strictly increasing for $\theta>0$. 

We consider a market agent that aims to sell a large amount of shares of a single stock with no time restrictions.
Let $Y_t$ denote the number of shares held by the agent at time 
$t$. We refer to a process $Y$ as a liquidation strategy if $Y_t$ tends to $0$ as $t$ tends to infinity. 
We consider the following set of admissible liquidation strategies. 
\begin{defi} \label{Def_adm}
For $y\in\mathbb{R}^+$, let $\mathcal{A}(y)$ denote the set of all
$(\mathcal{F}_t)_{t\geq 0}$-adapted, predictable, decreasing, \dg 
processes $Y$, satisfying $Y_{0-}=y$ and
\begin{gather}\label{stratAssump1}
\int_0^\infty \kappa_A\bigl(\Arrowvert Y_t\Arrowvert_{L^\infty(\ktP)}\bigr)\, dt<\infty .
\end{gather}
Moreover, let $\mathcal{A}_D(y)$ denote the set of all 
deterministic strategies in $\mathcal{A}(y)$.
\end{defi}
%\noindent

\begin{rem}
Condition (\ref{stratAssump1}) amounts to the following 
\begin{align*}
\int_0^\infty \Arrowvert Y_t\Arrowvert_{L^\infty(\mathbb{P})}\, dt&<\infty,\qquad\text{if }\mu<0 ,\\
\int_0^\infty \Arrowvert Y_t\Arrowvert_{L^\infty(\mathbb{P})}^2\, dt&<\infty,\qquad\text{if }\mu=0 .
\end{align*} 
When $\mu>0$ (which we do not consider), condition (\ref{stratAssump1}) would have to be
replaced with a condition like the one for $\mu<0$ above.
Conditions like these are not just technical conditions, but are necessary in order for the problem
that we consider in this paper to be meaningful. If we do not impose any restrictions, then the
value function is infinite. With reference to (\ref{opt}), the reason is that since there is no discounting, 
there is no penalty in
waiting until the best bid price reach a certain level and then start liquidating. For instance if
$L$ is a Brownian motion, then the time it takes for the best bid price to reach a given level $p$
is finite almost surely, for any level $p$. We skip the mathematical details, but intuitively 
a higher price level $p$ results in a higher
expected utility of the cash position. Consequently the value function will be infinite if such
strategies are admissible.
\end{rem}

To describe the agent's execution price, we explicitly model a bid limit order book. 
We assume that the unaffected bid price process $B^0$, which is the process describing the best 
bid prices in the market if the agent does not trade, is given by
\begin{gather*}
B_t^0=b+L_t ,\qquad t\geq 0 ,
\end{gather*}
where $b>0$ is the best bid price at the initial time. This Bachelier-L\'{e}vy type price model may seem simplistic, 
but this kind of modelling of the unaffected price process is widely used 
in the optimal liquidation literature \citep{AC2}, \citep{KM}, \citep{SS} and \citep{Gath}. 
There are studies which show that liquidation models with linear price processes 
provide a good approximation to models with exponential price processes 
\citep{GS} and \citet{FKTW}.

%In our model, the unaffected bid price is assumed to provide a lower bound for the best ask price 
%and that the best bid price as well as all bid prices are unaffected by the 
%large investor's buy orders (if he is allowed to buy back). 

%---------------------------------------
%NOTE!!!!!
%These assumptions are satisfied throughout 
%the whole chapter, and they allow us to exclude any buy orders 
%in the optimal trading strategy (see Remark \ref{Remnobuy}), and they also 
%exclude price manipulation in our model (see Remark \ref{Remnomani}). 

In order to describe the bid limit order book, we consider 
a measure $m$ defined on the Borel $\sigma$-algebra on $\ktR^-$, 
denoted by $\mathcal{B}(\ktR^-)$. If $\mathcal{S}\in\mathcal{B}(\ktR^-)$, 
then $m(\mathcal{S})$ represents the number of bid orders with prices 
in the set $B_t^0+\mathcal{S}=\{B_t^0+s\mid s\in \mathcal{S}\}$, provided 
that the agent did not make any trades before time $t$. Notice that the undisturbed 
bid order book described by $m$ is relative to the unaffected bid prices in the sense that it 
shifts together with the movement of the unaffected price. 
We impose the following assumptions on $m$.
\begin{assum}\label{Assumpmu}
We assume that 
\begin{itemize}
\item[(i)] there exists some $\bar{x}\in(-\infty,0)$ such that $m((\bar{x},0])=m(\ktR^-)<\infty$,
\item[(ii)] $m$ is absolutely continuous with respect to Lebesgue measure, 
and is non-zero on any interval properly containing the origin,
\item[(iii)] the function $x\mapsto m((x,0])$ is concave in $x$, for $x\in\mathbb{R}^-$.
%where $\mu'((x,0])$ is the derivative of $\mu((x,0])$ with respect to $x$. 
\end{itemize}
\end{assum}
%\noindent
%Note that the condition $\lim_{x\rightarrow-\infty}\frac{x}{\mu((x,0])}<0$ implies that 
%$\mu$ is $\sigma$-finite. 
%The first assumption means that there are finitely many bid orders available in the order book and the finite number $\bar{x}$ 
%is equal to the smallest bid price in the book. We know from (ii) that the right end of the bid order book 
%coincides with the best bid price in the undisturbed bid order book. In other words, one can always sell some 
%amount of shares at the unaffected bid price in an undisturbed bid order book. 
The concavity of $x\mapsto m((x,0])$ 
means that there are less and less bid orders in the undisturbed limit order book the further we get from the best bid price. 
Set $\bar{z}=-m(\ktR^-)$, which represents the total amount of bid limit orders in the undisturbed limit order book,
and introduce the functions $\phi:[-\infty,0]\rightarrow\mathbb{R}^-$ 
and $\psi:\mathbb{R}^-\rightarrow[-\infty,0]$ by 
\begin{gather*}
\phi(x)=-m\bigl((x,0]\bigr)\quad\text{ and }\quad
\psi(z)=\phi^{-1}(z), 
%\sup\bigl\{x\leq 0\mid \phi(x)\leq z\bigr\},
\end{gather*}
where $\phi(\psi(z))=z$, for all $z\in[\bar{z},0]$, and $\psi(z)=-\infty$, for all $z<\bar{z}$. 
A consequence of Assumption \ref{Assumpmu} is that $\phi$ is convex, $\psi$ is concave, 
and  $\phi$ and $\psi$ are both continuous and strictly increasing on their effective domain. They also satisfy 
\begin{align} \label{phipsicdt1}
\phi(0)=\psi(0)=0,
\end{align}
%\begin{align} \label{phipsicdt2}
%\lim_{x\rightarrow-\infty}\phi(x)=\bar{z};\,\,\, 
%%\text{ and }\,\,\,\lim_{z\rightarrow-\infty}\psi(z)=-\infty; 
%\end{align} 
%\begin{gather} \label{phipsicdt3}
%\lim_{x\rightarrow-\infty}\frac{x}{\phi(x)}=\infty\,\,\,\text{ and }\,\,\,
%\lim_{z\rightarrow\bar{z}}\frac{\psi(z)}{z}=\infty;  
%\end{gather}
as well as
\begin{gather} \label{phipsicdt4}
\int_0^{\bar{z}} \psi(u)\,du<\infty \quad\text{ and }\quad \psi(\bar{z})>-\infty.  
\end{gather}

%The state of the limit order book changes during trading. The book recovers 
%as new limit orders arrives. 
In order to model the dynamic of the bid order book during trading, we need to 
introduce one more process that captures the state of the order book. 
For a given strategy $Y$, let $Z^Y$ be an $\mathbb{R}^-$-valued process 
such that $-Z_t^Y$ represents the volume spread at time $t$. That is $-Z_t^Y$ is equal to the total number of bid orders which have 
already been executed subtracted by the total amount of limit orders which have arrived to refill the book up to time $t$. 
We call $Z^Y$ the state process of the bid limit order book associated with a trading strategy $Y$. 
Let $Z^Y_{0-}=z$, where $z\geq\bar{z}$ is the initial state of our bid order book. 
%\footnote{If $z<0$, then the bid order book is not in its undisturbed state at the initial time. This is interpreted as that there was some other large investor traded before our investor starts to trade.} 
Therefore, we have $\psi(Z_t^Y)=B_t^Y-B_t^0$, where $B_t^Y$ is the best bid price at time $t$ corresponding 
to $Y$, and $\psi(Z_t^Y)$ can be understood as the extra price spread at time $t$, caused by the investor 
who implements a strategy $Y$. Note that we have defined $\psi(z)=-\infty$, 
for all $z<\bar{z}$. This implies that the best bid price drops down to $-\infty$, if one sell more shares than 
available bids in the book. 
%This in particular will exclude the possibility that the investor making sale 
%while there is no available bid orders. 
The rate at which bid orders are refilling the order book is described by a resilience function 
$h:\ktR^-\rightarrow\ktR^-$ which satisfies the following. 
%\noindent
%%%%%%%%%%%%%%%%%%%%%%%%%%%%%%%%%%%%%%%%%%%%%%%%%%%%%%%%%%%%%%%%%%%%%%%%%%%%%%%%%%%%%%
%\begin{figure}
%\begin{picture} (150,52)
%\put(0,0){
%\begin{picture}(150,210) %\label{fig1}
%\thicklines

%\put(10,11){\vector(1,0){115}}    % x axis downleft
%\put(127,10){Price}
%\put(10,11){\vector(0,1){35}}    % y axis downleft
%\put(12,42){Number of Shares}

%\put(25,14){\line(1,0){16}}
%\put(41,19){\line(1,0){15}}
%\put(56,24){\line(1,0){14}}
%\put(70,29){\line(1,0){5}}
%\multiput(75,29)(2,0){4}{\line(1,0){1}}
%\multiput(83,34)(2,0){6}{\line(1,0){1}}
%\multiput(95,39)(2,0){6}{\line(1,0){1}}
%\put(83,34){\line(1,0){12}}
%\put(95,39){\line(1,0){11}}

%\put(25,11){\line(0,1){3}}
%\put(41,14){\line(0,1){5}}
%\put(56,19){\line(0,1){5}}
%\put(70,24){\line(0,1){5}}
%\multiput(83,29)(0,2){3}{\line(0,1){1}}
%\multiput(95,34)(0,2){3}{\line(0,1){1}}

%\multiput(106,37)(0,-2){14}{\line(0,1){1}}
%\put(115,11){\line(0,1){1.5}}

%\put(75,29){\line(0,-1){18}}

%\put(115,6.5){$b$}
%\put(104,6.5){$B^0_t$}
%\put(18,6.5){$B^0_t+\bar{x}$}
%\put(72,6.5){$B^Y_t$}
%\put(87,20){$-Z^Y_t$}
%\end{picture}
%}
%\put(10,1){\small{Figure 3.3: An illustration of a disturbed bid limit order book at time $t$ associated }}
%\put(10,-3){\small{with a strategy $Y$. }}
%\end{picture}
%\end{figure}
%%%%%%%%%%%%%%%%%%%%%%%%%%%%%%%%%%%%%%%%%%%%%%%%%%%%%%%%%%%%%%%%%%%%%%%%%%%%%%%%%%%%%%%%%

%\newpage
\begin{assum} \label{Assumh}
We assume that the resilience function $h:\ktR^-\rightarrow\ktR^-$
is increasing, locally Lipschitz continuous, satisfies $h(0)=0$ with $h(x)<0$ for all $x<0$,
and that the function
%and $\lim_{x\rightarrow-\infty}h(x)>-\infty$. 
$x\mapsto 1/h(x)$ is a concave function for $x<0$. 
\end{assum}
Note that the choice $h(x)=\lambda x$, for $\lambda>0$, which corresponds to exponential resilience,
satisfies Assumption \ref{Assumh}.
%\noindent
We then consider the state process $Z^Y$ with dynamic
\begin{gather}\label{Zeq}
dZ_t^Y=-h\bigl(Z_{t-}^Y\bigr)\, dt+ dY_{t} ,\qquad Z_{0-}^Y=z\in\ktR^-.
\end{gather}
For any admissible strategy $Y$, we refer to \citet{PSS} Appendix A for the existence and uniqueness 
of a negative, \dg and adapted solution to this dynamic. From Assumption \ref{Assumh} and equation 
(\ref{Zeq}) we observe that the further the best bid price is 
away from the unaffected bid price, the larger the speed of resilience for the best bid price. 

If the agent
does not  make any trades from time $t_1$ to $t_2$, then $\bigl(Z^Y_t\bigr)_{t_1<t<t_2}$ satisfies 
\begin{gather}
dZ^Y_t=-h(Z^Y_t)\, dt.  \label{dZ=-h(Z)dt}
\end{gather}
Now define a strictly decreasing function $H:\ktR^-\rightarrow\ktR\cup\{-\infty\}$ by 
\begin{gather}
H(x)=\int_{-1}^x\frac{1}{h(u)}\, du .  \label{Hdef}
\end{gather}
Let $H^{-1}$ denote the inverse of $H$, which satisfies $H^{-1}\bigl(H(x)\bigr)=x$ 
for all $x\leq 0$ and $H^{-1}(u)=0$ for $u\in\bigl(-\infty\,,\,\lim_{x\rightarrow 0-}H(x)\bigr]$. 
Then, it can be verified that the process $Z$ given by 
\begin{gather}
Z_t=H^{-1}\bigl(H(Z_0)-t\bigr)  \label{Zt-in-terms-of-H}
\end{gather}
has dynamic (\ref{dZ=-h(Z)dt}). 
Hence, for any $t$ between time $t_1$ and $t_2$, $Z_t^{Y}=H^{-1}\bigl(H\bigl(Z^Y_{t_1}\bigr)-t+t_1\bigr)$.
Moreover, if $Z^Y_{t_2}<0$, then 
\begin{gather}
t_2-t_1=H(Z^Y_{t_1})-H\bigl(Z^Y_{t_2}\bigr) .  \label{t_2-in-terms-of-H}
\end{gather}

Suppose that the agent's initial cash position is $c$ and that the agent 
implements a strategy $Y\in\mathcal{A}(y)$. Then the agent's cash position at time $T>0$ is
\begin{align}\label{cashnew1}
C_T(Y)=c-\int_0^T B_{t-}^Y\, dY_t^c -\sum_{0\leq t\leq T}\int_{0}^{\triangle Y_t}
\bigl\{B_{t-}^0+\psi\bigl(Z_{t-}^Y+x\bigr)\bigr\}\, dx ,
\end{align}
which corresponds to the best bids offered at all times being executed 
first so as to match the agent's orders, where the first integral 
represents the cost from the continuous component of the liquidation strategy and 
the sum of integrals represents the total cost due to all block sales. We also suppose 
the agent has a constant absolute risk aversion (CARA). With initial cash position 
$c$, an initial share position $y$ and infinite time-horizon, the agent wants to maximise 
the expected utility of the cash position at the end of time. Mathematically, 
the agent's optimal liquidation problem is
\begin{gather}\label{opt}
\sup_{Y\in\mathcal{A}(y)}\mathbb{E}\bigl[U\bigl(C_\infty(Y)\bigr)\bigr] ,
\end{gather}
where the utility function $U$ is given by
\begin{gather*}
U(c)=-\mathrm{e}^{-A c} , \qquad A>0 .
\end{gather*}
%This can be seen as a generalisation of the problem considered in
%\citet{Lokk}, and a risk-averse version of the problem considered
%by \citet{PSS}.

Observe that if $Z_t^Y<\bar{z}$, then $B_t^Y=B_t^0+\psi(Z_t^Y)=-\infty$. Clearly receiving the price $-\infty$
is unfavourable to the agent. Indeed, 
(\ref{cashnew1}) shows that this brings the agent an infinite cost. Due to this consideration, we will from 
now on only focus on admissible strategies $Y$ for which $Z_t^Y\geq\bar{z}$, for all $t\geq 0$. 
Define the function $\kappa_A:\mathbb{R}^+\rightarrow[0,\infty]$ by
\begin{gather*}
\kappa_A(y)=\kappa(-Ay) ,\qquad y\geq 0 ,
\end{gather*}
and set
\begin{gather*}
\bar{y}_A=\sup\bigl\{y\geq 0\mid \kappa_A(y)<\infty\bigr\} .
\end{gather*}
Then $\kappa_A$ is strictly increasing, strictly convex  and continuously differentiable on $[0,\bar{y}_A)$, 
with $\kappa_A(0)=0$. 
%According to Taylor's theorem, (\textcolor{red}{$\kappa''(y)$ may not exist!})
%\begin{gather}  \label{kappaA(y)-y-small}
%\kappa_A(y)=\frac{1}{2}\kappa_A''(y^*)y^2 ,\qquad\text{for some } y^*\in[0,y] . 
%\end{gather} 

%NOTE!!!
%With reference to (\ref{cumulant_func}), one can deduce that there exist
%$\epsilon, C_1, C_2>0$ such that
%\begin{gather}
%C_1y^2\leq \kappa_A(y)\leq C_2y^2 ,\qquad 0\leq y\leq\epsilon .  \label{kappaAp}
%\end{gather}
%\footnote{
%e.g.,choose $\epsilon\geq y$ such that 
%$C_1=\frac{1}{2}\inf_{y^*\in[0,\epsilon]}\kappa_A''(y^*)>0$ and 
%$C_2=\frac{1}{2}\sup_{y^*\in[0,\epsilon]}\kappa_A''(y^*)<\infty$}
%The function $\kappa_A$ will play a predominant role in the sequel.

%In our analysis, the assumption of constant absolute risk aversion
%is crucial in order to obtain explicit solutions.

%====================================================================
%====================================================================
%========== End of problem formulation ==============================
%====================================================================
%====================================================================

\section{Problem simplification}

In this section, we show that the utility maximisation problem in 
(\ref{opt}) can be reduced to a deterministic optimisation problem. This 
kind of result was first derived in \citet{SST}, which proved that with a 
certain market structure and an agent with constant absolute risk aversion, 
the optimal liquidation strategy is deterministic. 
%Some results of no price manipulation 
%strategies in our model will also be given in this section. 

Let $Y\in\mathcal{A}(y)$. Then
it follows from (\ref{cashnew1}) that
\begin{align*}
C_T(Y)=
c+by-(b+L_T)Y_T+\int_0^T Y_{t-} \,dL_t+\sum_{0\leq t\leq T}\triangle L_t\triangle Y_t-F_T(Y) ,
\end{align*}
where $F_T$ is given by
\begin{gather} \label{Fdef}
F_T(Y)=\int_0^T \psi\bigl(Z_{t-}^{Y}\bigr)\, dY_t^c
+\sum_{0\leq t\leq T}\int_0^{\triangle Y_t}
\psi\bigl(Z_{t-}^{Y}+x\bigr)\, dx .
\end{gather}
Since $t\mapsto \Arrowvert Y_T\Arrowvert_{L^\infty(\mathbb{P})}$ is decreasing, condition (\ref{stratAssump1}) implies 
that any admissible strategy $Y\in\mathcal{A}(y)$ satisfies 
\begin{gather}\label{tkappa}
\lim_{t\rightarrow\infty} t\,
\kappa_A\bigl(\Arrowvert Y_t\Arrowvert_{L^\infty(\mathbb{P})}\bigr)=0 .
\end{gather}
Also observe that
\begin{gather*}
\lim_{x\rightarrow 0} \frac{\kappa_A(x)}{x}=-A\mu .
\end{gather*}
Therefore, if $\mu<0$, there exists an $\epsilon>0$ and constants $C_1,C_2>0$ such that
\begin{gather*}
C_1x\leq\kappa_A(x)\leq C_2 x,\qquad\text{ for } x\in[0,\epsilon] . 
\end{gather*}
With reference to (\ref{tkappa}), It follows that for every $Y\in\mathcal{A}(y)$,
\begin{gather}\label{muneg}
\lim_{t\rightarrow\infty}t\Arrowvert Y_t\Arrowvert_{L^\infty(\mathbb{P})}=0 ,\qquad\text{ if } \mu<0 .
\end{gather}
If $\mu=0$, then
\begin{gather*}
\lim_{x\rightarrow 0} \frac{\kappa_A(x)}{x^2}=K ,\qquad\text{ for some }K>0 .
\end{gather*}
Therefore, if $\mu=0$, there exists an $\epsilon>0$ and constants $C_1,C_2>0$ such that
\begin{gather*}
C_1x^2\leq\kappa_A(x)\leq C_2 x^2,\qquad\text{ for } x\in[0,\epsilon] . 
\end{gather*}
With reference to (\ref{tkappa}) it follows that for every $Y\in\mathcal{A}(y)$,
\begin{gather}\label{munull}
\lim_{t\rightarrow\infty}t\Arrowvert Y_t\Arrowvert_{L^\infty(\mathbb{P})}^2=0 ,\qquad\text{ if } \mu=0 .
\end{gather}

Let $Y$ be an admissible strategy in $\mathcal{A}(y)$. Then with reference to (\ref{muneg}) and (\ref{munull}),
we calculate
\begin{gather*}
\lim_{T\rightarrow\infty}\mathbb{E}\bigl[\arrowvert L_TY_T\arrowvert^2\bigr]
\leq \lim_{T\rightarrow\infty}
\bigl(\mu^2T^2+
\kappa''(0)T \bigr)
\Arrowvert Y_T\Arrowvert_{L^\infty(\mathbb{P})}^2
=0 .
\end{gather*}
We conclude that $B_T^0Y_T$ tends to $0$ in $L^2(\mathbb{P})$
as $T\rightarrow\infty$. Furthermore, 
%set
%\begin{gather*}
%t_\epsilon=\inf\bigl\{t\geq 0\mid \Arrowvert Y_t\Arrowvert_{L^\infty(\mathbb{P})}\leq \epsilon\bigr\} .
%\end{gather*}
%Then
\begin{align*}
\mathbb{E}\biggl[\biggr(\int_0^\infty Y_{t-}\, dL_t\biggr)^2\biggr]^{\frac{1}{2}}
&\leq \mathbb{E}\biggl[\biggr(\int_0^\infty Y_{t-} \mu\, dt\biggr)^2\biggr]^{\frac{1}{2}}
+\mathbb{E}\biggl[\biggr(\int_0^\infty Y_{t-}\, d\bigl(L_t-\mu t\bigr)\biggr)^2\biggr]^{\frac{1}{2}}\cr
&\leq\mu \int_0^\infty \Arrowvert Y_t\Arrowvert_{L^\infty(\mathbb{P})}\, dt
+\biggl[\kappa''(0)\int_0^\infty \Arrowvert Y_t\Arrowvert_{L^\infty(\mathbb{P})}^2\, dt \biggr]^{\frac{1}{2}}
%&\leq \kappa''(0)\biggl(y^2 t_\epsilon 
%+\int_{t_\epsilon}^\infty
%\Arrowvert Y_t\Arrowvert_{L^\infty(\mathbb{P})}^2\, dt \biggr)\cr
%&\leq \kappa''(0)\biggl(y^2 t_\epsilon 
%+C_1^{-1}\int_{t_\epsilon}^\infty
%\kappa_A\bigl(\Arrowvert Y_t\Arrowvert_{L^\infty(\mathbb{P})}\bigr)\, dt\biggr)
<\infty .
\end{align*}
Hence, $\int_0^\infty Y_{t-}\, dL_t$ is well-defined in $L^2(\ktP)$. Due to the 
predictability of $Y$, we also have that 
\begin{gather*}
\mathbb{E}\biggl[\biggl(\sum_{0\leq t\leq T}\triangle L_t\triangle Y_t\biggr)^2\biggr]
=\mathbb{E}\biggl[\int_0^T\bigl(\triangle Y_t\bigr)^2\,dt\biggr]
\biggl(\int_{\mathbb{R}\setminus\{0\}} z^2\nu(dz)\biggr)=0, 
\end{gather*}
for all $T>0$, which shows that the quadratic covariation of the jumps of $L$ and $Y$ is almost surely 0. 
Moreover, note that $F_T(Y)\geq 0$ is an increasing 
function of $T$. Therefore, $F_\infty$ is a well defined function from the 
set of c\`{a}dl\`{a}g non-increasing functions into the extended positive 
real numbers. The final cash position is hence given by 
\begin{gather}
C_\infty(Y)=
c+by+\int_0^\infty Y_{t-} \,dL_t-F_\infty(Y), \label{Cinfinity}
\end{gather}
where $c+by$ represents the mark-to-market value of the total wealth of the agent's position 
at the start of the liquidation, $\int_0^\infty Y_{t-} \,dL_t$ represents the profit or loss 
due to the market risk, and  $F_\infty(Y)$ represents the cost due to the price impact.

Let $Y\in\mathcal{A}(y)$ and
define the process
$M^Y$ by
\begin{gather*}
M^Y_t=\exp\biggl(-A\int_0^t  Y_{s-}\, dL_s-\int_0^t
\kappa_A\bigl(Y_{s-}\bigr)\, ds\biggr), \qquad t\geq 0 .
\end{gather*}
Then it follows from Theorem 3.2 in \citet{KS} that $M^Y$ is a uniformly integrable martingale. 
%I.e. $M_t=\mathbb{E}\bigl[M_\infty\arrowvert\mathcal{F}_t\bigr]$,
%for all $t\geq 0$.
We can therefore define a probability measure $\widetilde{\mathbb{P}}=\mathbb{P}^Y$ by
\begin{gather*}
\frac{d\widetilde{\mathbb{P}}}{d\mathbb{P}}=M^Y_\infty .
\end{gather*}
Following the idea of the proof of Theorem 2.1 in \citet{SST}, we set
\begin{gather*}
I=\inf_{Y\in\mathcal{A}_D(y)}\int_0^\infty \kappa_A\bigl(Y_{t-}\bigr) \, dt
+A F_\infty(Y) ,
\end{gather*}
and note that $\kappa_A(\cdot)$ and $F_\infty(\cdot)$ are deterministic. Let
$Y^\epsilon\in\mathcal{A}_D(y)$ be such that
\begin{gather*}
\int_0^\infty \kappa_A\bigl(Y_{t-}^\epsilon\bigr) \, dt
+A F_\infty(Y^\epsilon) \leq I+\epsilon .
\end{gather*}
For an arbitrary $Y\in\mathcal{A}(y)$, we calculate that 
\begin{align*}
\mathbb{E}\bigl[U\bigl(C_\infty(Y)\bigr)\bigr]
&=-\mathrm{e}^{-A(c+by)}\mathbb{E}\biggl[
\exp\biggl(-A\int_0^\infty Y_{t-} \,dL_t+A F_\infty(Y)
\biggr)\biggr]\cr
&=-\mathrm{e}^{-A(c+by)}\mathbb{E}\biggl[
M_\infty\exp\biggl(\int_0^\infty \kappa_A \bigl(Y_{t-}\bigr) \, dt
+A F_\infty(Y)\biggr)\biggr]\cr
&= -\mathrm{e}^{-A(c+by)}\widetilde{\mathbb{E}}\biggl[
\exp\biggl(\int_0^\infty \kappa_A\bigl(Y_{t-}\bigr) \, dt
+A F_\infty(Y)\biggr)\biggr]\cr
&\leq -\mathrm{e}^{-A(c+by)}\mathrm{e}^{-\epsilon}
\widetilde{\mathbb{E}}\biggl[
\exp\biggl(\int_0^\infty \kappa_A\bigl(Y_{t-}^\epsilon\bigr) \, dt
+A F_\infty(Y^\epsilon)\biggr)\biggr]\cr
&\leq-\mathrm{e}^{-A(c+by)}\mathrm{e}^{-\epsilon}
\exp\biggl(\inf_{Y\in\mathcal{A}_D(y)}
\biggl\{\int_0^\infty \kappa_A\bigl(Y_{t-}\bigr) \, dt
+A F_\infty(Y)\biggr\}\biggr) ,
\end{align*}
since
\begin{gather*}
\int_0^\infty \kappa_A\bigl(Y_{t-}\bigr) \, dt
+A F_\infty(Y) \geq I\geq \int_0^\infty \kappa_A\bigl(Y_{t-}^\epsilon\bigr) \, dt
+A F_\infty(Y^\epsilon)-\epsilon . 
\end{gather*}
By letting $\epsilon$ tend to $0$ and taking the supremum over all admissible strategies on the left-hand side
of (\ref{red-eq}), we obtain
\begin{gather}\label{red-eq}
\sup_{Y\in\mathcal{A}(y)}\mathbb{E}\bigl[U\bigl(C_\infty(Y)\bigr)\bigr]
=-\mathrm{e}^{-A(c+by)}
\exp\biggl(\inf_{Y\in\mathcal{A}_D(y)}
\biggl\{\int_0^\infty \kappa_A\bigl(Y_{t-}\bigr) \, dt
+A F_\infty(Y)\biggr\}\biggr) 
\end{gather}

\begin{lem}\label{LemmC}
Let $F$ be given by (\ref{Fdef}). Then for every $Y\in\mathcal{A}_D(y)$ and $z\in[\bar{z},0]$, 
\begin{align}\label{Fexpr}
F_\infty(Y)=\int_{z}^{0}\psi(s)\, ds+
\int_0^\infty h\bigl(Z_{t-}^Y\bigr)\psi\bigl(Z_{t-}^Y\bigr)\, dt .
\end{align}
\end{lem}

With reference to Lemma \ref{LemmC} and (\ref{red-eq}), the optimal liquidation problem amounts to 
solving
\begin{gather}\label{Vdef2}
V(y,z)=\inf_{Y\in\mathcal{A}_D(y)}\int_0^\infty \biggl(
\kappa_A\bigl(Y_{t-}\bigr)+A h(Z_{t-}^Y)\psi(Z_{t-}^Y)\biggr)\, dt ,
\end{gather}
with $y=Y_{0-}$ and $z=Z_{0-}^Y$. Since $h$ and $\psi$ are both negative-valued and $\kappa_A\geq 0$, 
we have $V\geq 0$. Suppose $y>\bar{y}_A$, which is the upper bound for which $\kappa_A$ is finite 
($\bar{y}_A$ might be $+\infty$). 
In this case, the market agent will want to make an immediate block sale to bring the number of shares less than $\bar{y}_A$, 
since otherwise $Y$ does not satisfy (\ref{stratAssump1}) and $V(y,z)=\infty$. 
However, if the agent sell more than $z-\bar{z}$ number of shares, the value function $V(y,z)$ will be infinite. 
We therefore define the solvency region to be 
\begin{gather*}
\mathcal{D}=\bigl\{\,(y,z)\in\ktR^+\times[\bar{z},0]\,\,\big|\,\,z>y-\bar{y}_A+\bar{z}\,\bigr\},
\end{gather*}
and for the remainder of the paper we focus on this region.
For technical reasons, we do not consider $z=y-\bar{y}_A+\bar{z}$, as the value function may explode 
also along this line.

%===================================================================
%===================================================================
%========= End of problem simplification ===================
%===================================================================
%===================================================================

\section{Solution to the problem}

Our next aim is to derive a solution to the 
problem (\ref{Vdef2}). The derivation will be based on applying a 
time-change, and  the principle of dynamic programming. With 
reference to the results in \citet{Lokk} and the general theory 
of optimal control \citep{FS}, it is natural to guess 
that there exists a 
decreasing\footnote{When the volume spread is small, but the stock position is large, it seems intuitive 
to sell rapidly. On the other hand, if the volume spread is large, but the stock position is small, it seems 
intuitive to wait for the order book to recover. This motivates us to guess that the optimal intervention 
boundary is decreasing and separates the $(y,z)$ domain. }
c\`{a}gl\`{a}d function 
%(((\gd on the interval where it si finite)))%
$\beta=\beta^*:\mathbb{R}^+\rightarrow [\bar{z},0]$ which separates 
the $(y,z)$ domain into two different regions, a region where the agent 
makes an immediate block sale and another where the agent waits for the order book
to recover.
Let $\beta_*$ denote the \dg version of $\beta^*$, and set
\begin{align*}
\overline{\mathcal{S}}^\beta&=\bigl\{(y,z)\in\mathcal{D}
\mid z\geq\beta_*(y)\bigr\},\cr
\overline{\mathcal{W}}^\beta&=\bigl\{(y,z)\in\mathcal{D}
\mid z\leq\beta^*(y)\bigr\}\cup\bigl\{(y,z)\mid y=0\bigr\},\cr
\mathcal{G}^\beta&=\overline{\mathcal{S}}^\beta\cap\overline{\mathcal{W}}^\beta .
\end{align*}
$\overline{\mathcal{S}}^\beta$ represents the immediate sales region, 
$\overline{\mathcal{W}}^\beta$ the waiting region, and $\mathcal{G}^\beta$ 
is the graph of the intervention boundary $\beta$ and represents the
continuous sales region. For $y>0$, the Hamilton-Jacobi-Bellman 
equation corresponding to $V$ given by (\ref{Vdef2}) takes the form
\begin{align}\label{h1}
D_y^-v(y,z)+v_z(y,z)&=0 ,\quad\text{ for }(y,z)\in\overline{\mathcal{S}}^\beta ,\\
h(z) v_z(y,z)-\kappa_A(y)-Ah(z)\psi(z)&\leq 0 ,\quad\text{ for }
(y,z)\in\overline{\mathcal{S}}^\beta\setminus\mathcal{G}^\beta ,\label{h2}
\end{align}
and
\begin{align}\label{h3}
h(z) v_z(y,z)-\kappa_A(y)-Ah(z)\psi(z)&=0 ,\quad\text{ for }
(y,z)\in\overline{\mathcal{W}}^\beta ,\\
D_y^-v(y,z)+v_z(y,z)&\leq 0 ,\quad\text{ for }
(y,z)\in\overline{\mathcal{W}}^\beta\setminus\mathcal{G}^\beta  ,\label{h4}
\end{align}
with associated boundary condition $v(0,z)=A\int_0^z \psi(u)\,du$ for all $z\in[\bar{z},0]$,
where\footnote{The value function turns out to be continuously differentiable in $z$, but only 
continuous with a one-sided derivative in $y$ (see Proposition \ref{Propv}).}
\begin{gather*}
D_y^-v(y,z)=\lim_{\epsilon\rightarrow 0^-}\frac{1}{\epsilon}\biggl(v(y+\epsilon,z)-v(y,z)\biggr) .
\end{gather*}
%and 
%\begin{gather*}
%v_z(y,z)=\lim_{\epsilon\rightarrow 0^+}\frac{1}{\epsilon}\biggl(v(y,z+\epsilon)-v(y,z)\biggr) .
%\end{gather*}
%Observe that (\ref{h1}) and (\ref{h4}) imply that
%\begin{gather*}
%\max_{0\leq \triangle\leq y}\bigl\{v(y,z)-v(y-\triangle,z-\triangle)\bigr\}\leq 0 .
%\end{gather*}
The equations (\ref{h1})--(\ref{h4}) can be motivated as follows. When the 
market agent is trying to optimise over deterministic strategies, the agent basically has two options.
The agent can either sell a certain number $\triangle>0$ of shares or wait. Given a state 
$(y,z)$, it may or may not be optimal to sell $\triangle$ amount of shares, thus
\begin{gather*} %\label{hjb-inf1}
v(y,z)\leq v\bigl(y-\triangle, z-\triangle\bigr) ,
\end{gather*}
because the share position is decreased from $y$ to $y-\triangle$, due to 
$\triangle$ number of shares being sold, while at the same time the state of 
the bid order book changes from $z$ to $z-\triangle$. This inequality should 
hold for all $0<\triangle\leq y$, therefore
\begin{gather}\label{hjb-inf2}
\max_{0<\triangle\leq y}\bigl\{v(y,z)-v(y-\triangle,z-\triangle)\bigr\}\leq 0 .
\end{gather}
On the other hand, during a period of time $\triangle t>0$, it may or may 
not be optimal to wait, hence
\begin{align*}
v(y,z)&\leq v\bigl(y, Z_{\triangle t}\bigr)
+\int_0^{\triangle t}\biggl(\kappa_A(y)+Ah(Z_{u-})\psi\bigl(Z_{u-}\bigr)\biggr)\, du \cr
&=v(y,z)+\int_0^{\triangle t}\biggl(\kappa_A(y)+Ah(Z_{u-})\bigl(Z_{u-}\bigr)
-v_z\bigl(y, Z_{u-}\bigr) h(Z_{u-})\biggr)\, du ,
\end{align*}
where $dZ_{u}=-h(Z_u)\, du$, for $0\leq u\leq \triangle t$. Multiplying
the above inequality by $(\triangle t)^{-1}$ and sending $\triangle t$ 
to $0$, we obtain
\begin{gather}\label{hjb-inf4}
h(z) v_z(y,z)-\kappa_A(y)-Ah(z)\psi(z)\leq 0 .
\end{gather}
Since one of these strategies should be optimal, 
equality should hold in either (\ref{hjb-inf2}) or (\ref{hjb-inf4}). 
We therefore get
\begin{align*} %\label{hjb2}
\max\biggl\{&\max_{0<\triangle\leq y}\bigl\{v(y,z)-v(y-\triangle,z-\triangle)\bigr\}\,\,,\,\, 
h(z)v_z(y,z)-\kappa_A(y)-Ah(z)\psi(z)\biggr\}=0,
\end{align*}
from which (\ref{h1})--(\ref{h4}) follow. 

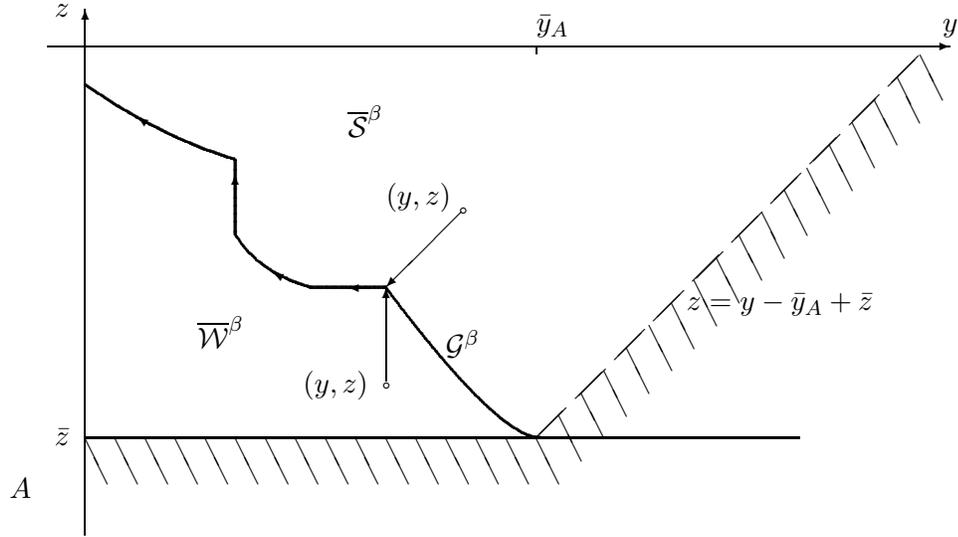
\begin{figure}
\begin{picture} (160,85)
\put(10,15){
\begin{picture}(120,95) %\label{fig1}
%\put(0,0){\line(1,0){120}} \put(120,0){\line(0,1){95}}
%\put(0,0){\line(0,1){95}} \put(0,95){\line(1,0){120}}   % borders
\put(5,60){\vector(1,0){120}}
\put(10,-5){\vector(0,1){70}}    % axes
\put(10,8){\line(1,0){95}}
\multiput(70,8)(5.2,5.2){10}{\line(1,1){4}}    %boundary
\multiput(10,7.9)(4,0){15}{\line(1,-2){3}}
\multiput(70,7.9)(3,3){18}{\line(1,-2){3}}
\put(0,0){\qbezier(10,55)(20,48)(30,45)} 		%h(y)
\put(0,-0.1){\qbezier(10,55)(20,48)(30,45)}
\put(0,0.1){\qbezier(10,55)(20,48)(30,45)}
\put(0,0){\qbezier(30,45)(30,40)(30,35)} 	
\put(-0.1,0){\qbezier(30,45)(30,40)(30,35)}
\put(0.1,0){\qbezier(30,45)(30,40)(30,35)}
\put(0,0){\qbezier(30,35)(33,30)(40,28)} 	
\put(0,-0.1){\qbezier(30,35)(33,30)(40,28)}
\put(0,0.1){\qbezier(30,35)(33,30)(40,28)} 
\put(0,0){\qbezier(40,28)(45,28)(50,28)} 	
\put(0,-0.1){\qbezier(40,28)(45,28)(50,28)}
\put(0,0.1){\qbezier(40,28)(45,28)(50,28)} 
\put(0,0){\qbezier(50,28)(65,8)(70.5,8)}
\put(0,-0.1){\qbezier(50,28)(65,8)(70.5,8)}
\put(0,0.1){\qbezier(50,28)(65,8)(70.5,8)} 
\put(70.5,7.9){\line(1,0){34.5}}
%\put(70.5,8.1){\line(1,0){34.5}}
\put(58,19){$\mathcal{G}^\beta$}
\put(45,48){$\overline{\mathcal{S}}^\beta$}
\put(25,20){$\overline{\mathcal{W}}^\beta$}
\put(6,7){$\bar{z}$}
\put(70,62){$\bar{y}_A$}
\put(90,25){$z=y-\bar{y}_A+\bar{z}$}
\put(70,60){\line(0,-1){1}}
\put(60,38){\vector(-1,-1){10}} \put(50,39){$(y,z)$} \put(60.3,38.2){\circle{0.8}}
\put(50,15.5){\vector(0,1){12.5}} \put(39,14){$(y,z)$} \put(50.05,15){\circle{0.8}}
\put(20,48.7){\vector(-3,2){3}}
\put(38,28.5){\vector(-2,1){3}}
%\put(31.8,53.1){\vector(-1,1){2}}
\put(29.95,40){\vector(0,1){3}}
\put(48,27.92){\vector(-1,0){3}}
\put(6,64){$z$} \put(124,62){$y$}
\put(0,0){$A$}
\end{picture}
}
%\put(17,0){\small{An illustration of the
%strategy $Y^\beta$ corresponding to a boundary $\beta$.}}
\end{picture}
\caption{An illustratation of the strategy $Y^\beta$ corresponding to an intervention boundary $\beta$
(the graph of $\beta$ is $\mathcal{G}^\beta$).
The solvency region $\mathcal{D}$ is the region that is not shaded. For an initial state 
$(y,z)\in\overline{\mathcal{S}}^\beta$, the strategy $Y^\beta$ consists of an initial block sale that brings
the state onto $\mathcal{G}^\beta$.
For an initial state 
$(y,z)\in\overline{\mathcal{W}}^\beta$, the strategy $Y^\beta$ consists of waiting while the order book recovers
until the state reach $\mathcal{G}^\beta$. Once the state is on $\mathcal{G}^\beta$, 
the strategy $Y^\beta$ consists of continuously submitting sales orders in such a way that the state
stays on $\mathcal{G}^\beta$.}
\label{betastrategyfig}
\end{figure}

We define the liquidation strategy $Y^\beta$ 
corresponding to an intervention boundary $\beta$ as the \dg function with the following properties:
\begin{itemize}
\item[(i)] If $(y,z)\in\overline{\mathcal{S}}^\beta$, then the agent  
makes an immediate block trade of size $\triangle$ such that
$(Y_0^\beta,Z_0^{Y^\beta})=(y-\triangle,z-\triangle)\in\mathcal{G}^\beta$, and set $t_w=0$.
\item[(ii)] If $(y,z)\in\overline{\mathcal{W}}^\beta$, then the agent waits until the 
time $t_w=\inf\bigl\{t\geq 0\mid Z_t^{Y^\beta}=\beta(y)\bigr\}$, where 
\begin{gather*}
Z_t^{Y^\beta}=z-\int_0^t h\bigl(Z_u^{Y^\beta}\bigr)\, du ,\qquad 0\leq t\leq t_w.
\end{gather*}
\item[(iii)]
For $t\geq t_w$, the agent continuously sell shares in such a way that
$(Y_t^\beta,Z_t^{Y^\beta})\in\mathcal{G}^\beta$, where
\begin{gather*}
Z_t^{Y^\beta}=Z_{t_w}^{Y^\beta}-\int_{t_w}^t h\bigl(Z_u^{Y^\beta}\bigr)\, du+Y_t^\beta-Y_{t_w}^\beta ,
\qquad t\geq t_w.
\end{gather*}
\item[(iv)] The agent takes no further action once $Y^\beta_t=0$.
\end{itemize}
Figure \ref{betastrategyfig} provides an illustration of such a strategy. 
We will later characterise an optimal intervention boundary, and prove that the corresponding strategy exists, 
is admissible and optimal. The key to characterise the optimal intervention boundary is that we are able to obtain
expressions for the performance of the strategy corresponding to a given intervention boundary.
% % % % % % % % % % % % % % % % % % % % % % % % % % % % % % % % % % % % % % % % % % % % % % % %
% % % % % % % % % % % % % % % % % % % % % % % % % % % % % % % % % % % % % % % % % % % % % % % % % % % %
But first, let us examine in more detail the strategy corresponding to a given intervention 
boundary function $\beta$. We will consider any intervention boundary $\beta:\mathbb{R}^+\rightarrow[\bar{z},0]$ 
which is decreasing, \gd  and satisfy
$\beta(y)<0$, for all $y>0$, $\lim_{y\rightarrow \infty}\beta(y)=\bar{z}$ 
and $\beta(0)=0$. 
%It will be shown later that there exists such an optimal intervention boundary 
%which completely characterises the solution to the investor's optimisation problem, 
%and the properties that the optimal boundary may be discontinuous (there might be countably 
%many discontinuities) and not invertible will complicate our analysis quite a lot. 
Now, given an intervention boundary $\beta$, one may ask whether the corresponding liquidation 
strategy $Y^\beta$ exists and is unique. In order to answer this, we need to introduce the 
following functions related to $\beta$.
\begin{align}\label{gammah}
\gamma_\beta(y)&=\beta(y)-y , \qquad\,\, \text{ for } y\in\mathbb{R}^+ ,\\
\rho_\beta(z)&=z-\beta^{-1}(z) ,\quad \ \text{ for } z\in[\bar{z},0].\label{rhoh}
\end{align}
We also introduce the inverse functions
\begin{align}
\beta^{-1}(z)&=\inf\bigl\{y\in\mathbb{R}^+\, \big|\, \beta(y)\leq z\bigr\}, 
\qquad\,\,\,\text{ for } z\in[\bar{z},0];\label{hinv}\\
\gamma_\beta^{-1}(x)&=\inf\bigl\{y\in\mathbb{R}^+\, \big|\, \gamma_\beta(y)\leq x\bigr\},
\qquad\text{ for } x\in\ktR^-;\label{gammahinv}\\
\rho_\beta^{-1}(x)&=\inf\bigl\{z\in[\bar{z},0]\, \big|\, \rho_\beta(z)\geq x\bigr\},
\quad\,\,\text{ for } x\in\ktR^-.\label{rhohinv}
\end{align}
Note that $\beta$ and $\gamma_\beta$ are \gd, $\beta^{-1}$ and
$\rho_\beta$ are \dg, and $\gamma_\beta^{-1}$ as well as $\rho_\beta^{-1}$ are
continuous\footnote{It can be checked that for $x\in\ktR^-$, $\gamma^{-1}_{\beta}(x)$ and $\rho^{-1}_{\beta}(x)$ 
is respectively the $y$-coordinate and the $z$-coordinate of the intersection of the line $z=y+x$ and 
$\mathcal{G}^\beta$.}.
Moreover, $\beta$, $\beta^{-1}$ and $\gamma_\beta^{-1}$ are decreasing, $\gamma_\beta$  
is strictly decreasing, $\rho_\beta$ is strictly increasing, and $\rho_\beta^{-1}$ is 
increasing. Furthermore, it follows directly from the definitions of $\beta^{-1}$, 
$\gamma_\beta$, $\gamma_\beta^{-1}$, $\rho_\beta$ and $\rho_\beta^{-1}$ that 
the following three identities hold. 
\begin{align}
\rho_\beta^{-1}(x)=x+\gamma_\beta^{-1}(x), &\qquad\text{ for all }x\in\ktR^-,\label{iden1}\\
\gamma_\beta^{-1}\bigl(\rho_\beta(z)\bigr)=\beta^{-1}(z), &\qquad\text{ for all }z\in[\bar{z},0], \label{iden2}\\
\rho_\beta^{-1}\bigl(\gamma_\beta(y)\bigr)=\beta(y), &\qquad\text{ for all }y\in\ktR^+. \label{iden3}
\end{align}
Also, by the definitions of $\mathcal{G}^\beta$, $\beta$ and $\beta^{-1}$, we 
see that the set $\mathcal{G}^\beta$ is the union of the graphs of the functions $\beta$ and $\beta^{-1}$
restricted to $\mathcal{D}$. 

Observe that if $z>\beta(y)$ , then the strategy $Y^\beta$ 
corresponding to the intervention boundary described by $\beta$ 
consists of an initial sale of $\triangle$ number of 
shares so that $(y-\triangle, z-\triangle)$ is in $\htG^\beta$ 
(see Figure \ref{betastrategyfig}). Let $Y_{0-}^\beta=y$ and $Y_0^\beta=y-\triangle$. 
Suppose $(y-\triangle, z-\triangle)$ is on the graph of $\beta$. 
Then $(y-\triangle, z-\triangle)=(y-\triangle, \beta(y-\triangle))$ 
and this equality is equivalent to
\begin{gather*}
\gamma_\beta\bigl(Y_0^\beta\bigr)=\beta(Y_0^\beta)-Y_0^\beta=z-y ,
\end{gather*}
from which it follows that $Y_0^\beta=\gamma_\beta^{-1}(z-y)$ 
and $\triangle=y-\gamma_\beta^{-1}(z-y)$. Now suppose 
$(y-\triangle, z-\triangle)$ is on the graph of $\beta^{-1}$, 
and let $Z_{0-}^{Y^\beta}=z$ and $Z_0^{Y^\beta}=z-\triangle$. 
Then $(y-\triangle, z-\triangle)=(\beta^{-1}(z-\triangle), z-\triangle)$, 
which is equivalent to
\begin{gather*}
\rho_\beta\bigl(Z_0^{Y^\beta}\bigr)=Z_0^{Y^\beta}-\beta^{-1}(Z_0^{Y^\beta})=z-y ,
\end{gather*}
and it follows that $Z_0^{Y^\beta}=\rho_\beta^{-1}(z-y)$ and 
$\triangle=z-\rho_\beta^{-1}(z-y)$. According (\ref{iden1}), the number 
$\triangle$ of shares in both of the aforementioned two cases can be expressed by 
\begin{gather*}
\triangle=y-\gamma_\beta^{-1}(z-y)=z-\rho_\beta^{-1}(z-y).
\end{gather*}
On the other hand, if $z\leq\beta(y)$, then the strategy $Y^\beta$ consists of an 
initial waiting period until $\bigl(Y_t^\beta,Z_t^{Y^\beta}\bigr)$ is on the graph of 
$\beta$ (see Figure \ref{betastrategyfig}). As long as no action is taken, we have $Y_t^\beta=y$, 
and with reference to (\ref{Zt-in-terms-of-H}) and (\ref{t_2-in-terms-of-H}), we 
obtain $Z_t^{Y^\beta}=H^{-1}\bigl(H(z)-t\bigr)$. The first time $t_w$ that the 
state process is on the graph of $\beta$ is therefore given by
\begin{gather}
t_w=H(z)-H\bigl(\beta(y)\bigr) .  \label{t_w-in-terms-of-H}
\end{gather}
Once the state process $(Y^\beta,Z^{Y^\beta})$ is in the set $\htG^\beta$,
the strategy $Y^\beta$ consists of selling shares in such
a way that the state process remains in $\htG^\beta$ (see Figure \ref{betastrategyfig}). 
Therefore, $\bigl(Y_t^\beta,Z_t^{Y^\beta}\bigr)=\bigl(Y_t^\beta, \beta(Y_t^\beta)\bigr)$ 
whenever $\beta\bigl(Y_t^\beta+\bigr)=\beta\bigl(Y_t^\beta\bigr)$. With reference to (\ref{Zeq}), 
this implies that $Y_t^\beta$ should solve
\begin{gather*}
d\beta\bigl(Y_t^\beta\bigr)=-h\bigl(\beta\bigl(Y_t^\beta\bigr)\bigr)\, dt+dY_t^\beta,
\end{gather*}
which is equivalent to
\begin{gather*}
d\gamma_\beta\bigl(Y_t^\beta\bigr)=-h\bigl(\beta\bigl(Y_t^\beta\bigr)\bigr)\, dt .
\end{gather*}
If $\beta^{-1}\bigl(Z_t^{Y^\beta}\bigr)=\beta^{-1}\bigl(Z_t^{Y^\beta}-\bigr)$, then 
$\bigl(Y_t^\beta,Z_t^{Y^\beta}\bigr)=\bigl(\beta^{-1}(Z_t^{Y^\beta}), Z_t^{Y^\beta}\bigr)$.
According to (\ref{Zeq}) and the definition of $\beta^{-1}$, $Z^{Y^\beta}$ should solve
\begin{gather*}
dZ_t^{Y^\beta}=-h\bigl(Z_{t-}^{Y^\beta}\bigr)\, dt.
\end{gather*}
%Now let's write $\mathcal{S}=\{y\in\ktR^+\,|\,\beta(y)>-\infty\}$. 
Set
\begin{gather}\label{twdef}
t_w=
\begin{cases}
0 ,\qquad&\text{ if }z>\beta(y),\cr
H(z)-H\bigl(\beta(y)\bigr),
\qquad&\text{ if } z\leq \beta(y),
\end{cases}
\end{gather}
and
\begin{gather}
\bar{t}=\inf\{t\geq 0\mid Y^\beta_t=0\}.  \label{tbar}
\end{gather}
Denote by $\{y_n\}_{n\in\mathbb{I}}$ the set of discontinuity points of $\beta$. Then $\mathbb{I}$ 
is countable since $\beta$ is c\`{a}gl\`{a}d. Define $\{t_n\}_{n\in\mathbb{I}}$ by
\begin{gather}
t_n=\inf\bigl\{t\geq t_w\mid Y^\beta_t=y_n\bigr\} ,  \label{tn}
\end{gather}
and $\{s_n\}_{n\in\mathbb{I}}$ by
\begin{gather}
s_n=\inf\bigl\{t\geq t_w\mid Y^\beta_t<y_n\bigr\} .  \label{sn}
\end{gather}
If $\{t\geq t_w\mid Y^\beta_t=y_n\bigr\}=\emptyset$, set $t_n=\infty$; and 
set $s_n=\infty$, if $\{t\geq t_w\mid Y^\beta_t<y_n\bigr\}=\emptyset$. 
The following result establish existence and uniqueness of such
a strategy $Y^\beta$ corresponding to a given intervention boundary $\beta$.

%=====================================================================
%=====================================================================
%============= Strategy Lemma ========================================
%=====================================================================
%=====================================================================

\begin{lem}\label{Lemmstrat}
Let $(y,z)\in\mathcal{D}$ where $y>0$, and let $\beta$ be an intervention boundary function. 
%Suppose $h$ is 
%a resilience function satisfying Assumption \ref{Assumh}, and 
Let $H$, $\beta^{-1}$, $\gamma_\beta$, $\gamma_\beta^{-1}$, $t_w$, $\bar{t}$, $y_n$, $t_n$ 
and $s_n$ be given by (\ref{Hdef}), (\ref{hinv}), (\ref{gammah}), (\ref{gammahinv}) 
and (\ref{twdef})--(\ref{sn}), respectively. Set 
$\bigl(Y^\beta_t\bigr)_{t\geq 0}=\bigl(Y^\beta_{t\wedge\bar{t}}\bigr)_{t\geq 0}$, 
with $Y^\beta_{0-}=y$, which denotes the decreasing \dg liquidation 
strategy corresponding to $\beta$, and  let $\bigl(Z_t^{Y^\beta}\bigr)_{t\geq 0}$, with 
$Z^{Y^\beta}_{0-}=z$, be the state process of the bid order book 
associated with $Y^\beta$. Suppose $Y^\beta$ satisfies the following description:
\begin{itemize}
%\item[(i)] If $y=0$, then liquidation is completed immediately; otherwise,
\item[(i)] If $z>\beta(y)$, 
           \begin{itemize}
                \item[(a)] when $y\in
                           \cup_{n\in\mathbb{I}}\bigl(z-\beta(y_n)+y_n\,,\,z-\beta(y_n+)+y_n\bigr]$, 
                           immediately sell $y-\gamma_\beta^{-1}(z-y)$ number of shares. 
                           This block trade ensures $Y_0^\beta=\beta^{-1}\bigl(Z_0^{Y^\beta}\bigr)$.
                \item[(b)] when $y\in \bigl(z,\infty\bigr)\,\setminus\,
                           \cup_{n\in\mathbb{I}}\bigl(z-\beta(y_n)+y_n\,,\,z-\beta(y_n+)+y_n\bigr]$, 
                           immediately sell $y-\gamma_\beta^{-1}(z-y)$ number of shares. 
                           This block trade ensures $Z_0^{Y^\beta}=\beta\bigl(Y_0^\beta\bigr)$.
           \end{itemize}
            Then continuously sell shares so that $\bigl(Y_{t}^\beta,Z_{t}^{Y^\beta}\bigr)\in\htG^\beta$ 
            for all $t\in [\,t_w,\bar{t}\,]$. 
\item[(ii)] If $z\leq\beta(y)$, then wait until time $t_w$. The time $t_w$ has the 
             property that $Z_{t_w}^{Y^\beta}=\beta(y)$. Continuously sell shares so that 
             $\bigl(Y_{t}^\beta,Z_{t}^{Y^\beta}\bigr)\in\htG^\beta$ for all $t\in [\,t_w,\bar{t}\,]$. 
\end{itemize}
Such a strategy $Y^\beta$ exists and is unique, and it is continuous for all $t>0$. In particular, 
\begin{gather}
Y_t^{\beta}=y_n
\quad\text{ for } t\in[\,t_w, \bar{t}\,]\cap\,\cup_{ n\in \mathbb{I}}[t_n,s_n),\label{Y=beta^-1(Z)}
\end{gather}
with corresponding $Z^{Y^\beta}_t$ being the unique solution to
\begin{gather}
dZ_t^{Y^\beta}=-h\bigl(Z_{t}^{Y^\beta}\bigr)\, dt,  \label{dynamic-of-Z}
\end{gather}
where 
\begin{align}
Z_{t_w}^{Y^\beta}=\rho_\beta^{-1}(z-y)\,\text{ if } z>\beta(y),\quad
\text{ and }\quad Z_{t_n}^{Y^\beta}=\beta\bigl(Y_{t_n-}^\beta\bigr)\,\text{ for } t_n>t_w.  \label{Z-initial}
\end{align}
%Note that in the case of $z\leq \beta(y)$, $t_w\notin [t_n,s_n)$. 
Moreover, 
\begin{gather}
Z_t^{Y^\beta}=\beta\bigl(Y_t^\beta\bigr) ,
\quad\text{ for } t\in[\,t_w, \bar{t}\,]\setminus\,\cup_{ n\in \mathbb{I}}[t_n,s_n),  \label{Z=beta(Y)}
\end{gather}
where $Y^\beta$ is the unique solution to
\begin{gather}
d\gamma_\beta\bigl(Y_t^\beta\bigr)=-h\bigl(\beta\bigl(Y_t^\beta\bigr)\bigr)\, dt,  \label{dynamic-of-Y}
\end{gather}
with
\begin{align}
Y_{t_w}^\beta=y \text{ if } z\leq \beta(y), \quad
Y_{t_w}^\beta=\gamma_\beta^{-1}(z-y)  \text{ if } z>\beta(y),\quad
\text{and}\quad Y_{s_n}^\beta=y_n\text{ for } s_n>t_w.  \label{Y-initial}
\end{align}
If $t_w>0$, then $Y_t^\beta=y$ and $Z_t^{Y^\beta}=H^{-1}\bigl(H(z)-t\bigr)$,
for $0\leq t\leq t_w$. 
\end{lem}
%\noindent
We can also describe $Z_t^{Y^\beta}$ for 
$t\in[\bar{t},\infty)$ since it satisfies (\ref{dynamic-of-Z}) with initial condition
\begin{gather}  \label{Ztbar-initial}
Z_{\bar{t}}^{Y^\beta}=
\begin{cases}
Z_{t_w}^{Y^\beta},\, &\text{ if } \bar{t}=t_w, \cr
z,\,  &\text{ if } \bar{t}<t_w, \cr
\beta(0+),\,  &\text{ if } \bar{t}>t_w.
\end{cases}
\end{gather}
The value $\beta(0+)$ can then be used to determine whether the liquidation period is finite
or not. More specifically, 
we have that $\beta(0+)<0$ implies $\bar{t}<\infty$. To see this, it is enough to consider 
\begin{gather*}
\gamma_\beta\bigl(Y^\beta_{\bar{t}}\bigr)-\gamma_\beta\bigl(Y^\beta_t\bigr)
=\int_t^{\bar{t}}-h\bigl(\beta\bigl(Y_u^\beta\bigr)\bigr)\, du
\end{gather*}
which follows from (\ref{dynamic-of-Y}) when there is no waiting period between the times $t$ 
and $\bar{t}$. To get a contradiction, suppose $\bar{t}=\infty$. Then it is clear that 
$\int_t^{\bar{t}}-h\bigl(\beta\bigl(Y_u^\beta\bigr)\bigr)\, du=\infty$, as 
$\beta\bigl(Y_u^\beta\bigr)$ is bounded away from 0 on the interval $(t,\bar{t})$. However, 
$\gamma_\beta\bigl(Y^\beta_{\bar{t}}\bigr)-\gamma_\beta\bigl(Y^\beta_t\bigr)$ is finite, so we get a
contradiction.
%((Note that in the above Lemma, if there exists some $\tilde{y}>0$ 
%such that $\beta(\tilde{y})=0$, then $Y^\beta$ will be trapped in 
%the largest$\tilde{y}>0$ such that $\beta(\tilde{y})=0$ and hence 
%$\bar{t}=\infty$. However, this kind of $\beta$ will be excluded 
%by Lemma \ref{LemGamma}. If for all $y>0$, $\beta(y)<0$, then 
%$Y^\beta_t$ converges to 0.))

It follows from the dynamics of $Z_{t}^{Y^\beta}$ that $Z^{Y^\beta}$ is \dg and increasing to 0. 
Moreover, the continuity of $Y^\beta_t$ for $t>0$ implies that $Z^{Y^\beta}$ 
is also continuous for all $t>0$. 

We now progress by deriving an explicit expression for the performance 
function associated with the strategy $Y^\beta$ described by Lemma 
\ref{Lemmstrat} for an arbitrary intervention boundary $\beta$. 
This expression can then later be used to derive
an explicit expression for the value function of our 
problem. For the strategy $Y^\beta$ with associated 
state process $Z^{Y^\beta}$, given an initial state $(y,z)$, and with reference 
to (\ref{Vdef2}), we define the performance function $J_\beta$ by
\begin{gather}\label{Jdef}
J_\beta(y,z)=\int_0^\infty \biggl(\kappa_A\bigl(Y_{t}^\beta\bigr)+A h(Z_{t}^{Y^\beta})
\psi\bigl(Z_{t}^{Y^\beta}\bigr)\biggr)\, dt ,
\end{gather}
%((NOT NECESSARILY FINITE, SINCE HAVEN'T PROVED ADMISSIBILITY))
where $Y_{0-}^{\beta}=y$, $Z_{0-}^{Y^\beta}=z$ and $(y,z)\in\mathcal{D}$. Since $\kappa_A(0)=0$, it follows that 
\begin{gather} 
\int_{\bar{t}}^\infty \biggl(\kappa_A\bigl(Y_{t}^\beta\bigr)+A h(Z_{t}^{Y^\beta})
\psi\bigl(Z_{t}^{Y^\beta}\bigr)\biggr)\, dt=A\int_0^{Z^{Y^\beta}_{\bar{t}}} \psi(u) \,du. \label{tbar-infty=0}
\end{gather}
Therefore, %in cases (i) of Lemma \ref{Lemmstrat}, 
\begin{gather} \label{JI(i)}
J_\beta(0,z)=A\int_0^z \psi(u) \,du. 
\end{gather}
\begin{lem}  \label{Lemmtw-infty-rewite}
Let $\beta$, $Y^\beta$, $Z^{Y^\beta}$, $t_w$ and $\bar{t}$ be 
defined as in Lemma \ref{Lemmstrat}. If $t_w<\bar{t}$, then
\begin{align*}
&\quad\,\int_{t_w}^{\infty} \biggl(\kappa_A\bigl(Y_{t}^\beta\bigr)+A h\bigl(Z_{t}^{Y^\beta}\bigr)
\psi\bigl(Z_{t}^{Y^\beta}\bigr)\biggr)\, dt\cr
&=\int_{\beta(0+)}^{Z_{t_w}^{Y^\beta}-Y^\beta_{t_w}}
\biggl(\frac{\kappa_A\bigl(\gamma_\beta^{-1}(u)\bigr)}{h\bigl(\rho_\beta^{-1}(u)\bigr)}
+A\psi\bigl(\rho_\beta^{-1}(u)\bigr)\biggr)\, du+A\int_0^{\beta(0+)} \psi(u) \,du, 
\end{align*}
where $\gamma_\beta^{-1}$ and $\rho_\beta^{-1}$ are defined by (\ref{gammahinv}) and (\ref{rhohinv}), respectively. 
\end{lem}
In case (i) (a) of Lemma \ref{Lemmstrat}, the strategy $Y^\beta$ 
consists of an initial sale of $y-\gamma_\beta^{-1}(z-y)=z-\rho_\beta^{-1}(z-y)$ 
number of shares. The state after the block sale is
$\bigl(Y_0^\beta,Z_0^{Y^\beta}\bigr)=\bigl(\beta^{-1}\bigl(\rho_\beta^{-1}(z-y)\bigr), \rho_\beta^{-1}(z-y)\bigr)$,
Hence, according to (\ref{Jdef}) and Lemma \ref{Lemmtw-infty-rewite}, 
\begin{align*} %\label{JI(ii)(b)}
J_\beta(y,z)&=J_\beta\Bigl(\beta^{-1}\bigl(\rho_\beta^{-1}(z-y)\bigr)\,,\, \rho_\beta^{-1}(z-y)\Bigr)\cr
%&=\int_{\beta(0+)}^{Z_{0}^{Y^\beta}-Y^\beta_{0}}
%\biggl(\frac{\kappa_A\bigl(\gamma_\beta^{-1}(u)\bigr)}{h\bigl(\rho_\beta^{-1}(u)\bigr)}
%+A\psi\bigl(\rho_\beta^{-1}(u)\bigr)-A\psi(u)\biggr)\, du\cr
&=\int_{\beta(0+)}^{\rho_\beta(\rho_\beta^{-1}(z-y))}
\biggl(\frac{\kappa_A\bigl(\gamma_\beta^{-1}(u)\bigr)}{h\bigl(\rho_\beta^{-1}(u)\bigr)}
+A\psi\bigl(\rho_\beta^{-1}(u)\bigr)\biggr)\, du+A\int_0^{\beta(0+)} \psi(u) \,du\cr
&=\int_{\beta(0+)}^{z-y}
\biggl(\frac{\kappa_A\bigl(\gamma_\beta^{-1}(u)\bigr)}{h\bigl(\rho_\beta^{-1}(u)\bigr)}
+A\psi\bigl(\rho_\beta^{-1}(u)\bigr)\biggr)\, du+A\int_0^{\beta(0+)} \psi(u) \,du. 
\end{align*}
In case (i) (b), we immediately sell $y-\gamma_\beta^{-1}(z-y)$ 
number of shares at the beginning. The state after the block sale 
is $\bigl(Y_0^\beta,Z_0^{Y^\beta}\bigr)=\bigl(\gamma^{-1}_\beta(z-y),\beta\bigl(\gamma^{-1}_\beta(z-y)\bigr)\bigr)$.
Hence, similar to the above calculation, we have
\begin{align*} %\label{JI(ii)(c)}
J_\beta(y,z)&=J_\beta\Bigl(\gamma^{-1}_\beta(z-y),\beta\bigl(\gamma^{-1}_\beta(z-y)\bigr)\Bigr)\cr
%&=\int_{\beta(0+)}^{Z_{0}^{Y^\beta}-Y^\beta_{0}}
%\biggl(\frac{\kappa_A\bigl(\gamma_\beta^{-1}(u)\bigr)}{h\bigl(\rho_\beta^{-1}(u)\bigr)}
%+A\psi\bigl(\rho_\beta^{-1}(u)\bigr)-A\psi(u)\biggr)\, du\cr
%&=\int_{\beta(0+)}^{\gamma_\beta(\gamma_\beta^{-1}(z-y))}
%\biggl(\frac{\kappa_A\bigl(\gamma_\beta^{-1}(u)\bigr)}{h\bigl(\rho_\beta^{-1}(u)\bigr)}
%+A\psi\bigl(\rho_\beta^{-1}(u)\bigr)-A\psi(u)\biggr)\, du\cr
&=\int_{\beta(0+)}^{z-y}
\biggl(\frac{\kappa_A\bigl(\gamma_\beta^{-1}(u)\bigr)}{h\bigl(\rho_\beta^{-1}(u)\bigr)}
+A\psi\bigl(\rho_\beta^{-1}(u)\bigr)\biggr)\, du+A\int_0^{\beta(0+)} \psi(u) \,du.
\end{align*}
We therefore conclude that for case (i) of Lemma \ref{Lemmstrat}, 
\begin{gather}
J_\beta(y,z)=\int_{\beta(0+)}^{z-y}
\biggl(\frac{\kappa_A\bigl(\gamma_\beta^{-1}(u)\bigr)}{h\bigl(\rho_\beta^{-1}(u)\bigr)}
+A\psi\bigl(\rho_\beta^{-1}(u)\bigr)\biggr)\, du+A\int_0^{\beta(0+)} \psi(u) \,du.  \label{JI(ii)(b)(c)}
\end{gather}
%((NOT NECESSARILY FINITE AT $\bar{z}$. INFINITE AT $z-y$ IS EXCLUDED BY Lemma \ref{LemGamma}, 
%SINCE Lemma \ref{Lemmtw-infty-rewite} AND $\beta(Y^\beta_0)>-\infty$))

Moreover, in case (ii), $z\leq\beta(y)$. So we need to wait until time $t_w>0$ at which 
$Z^{Y^\beta}_{t_w}=\beta(y)$. With reference to (\ref{Hdef}) and (\ref{t_w-in-terms-of-H}), we have 
\begin{gather*}
t_w=H(z)-H\bigl(\beta(y)\bigr)=\int_{\beta(y)}^z\frac{1}{h(u)}\, du . 
\end{gather*}
Also, observe that
\begin{align*}
\int_0^{t_w}h\bigl(Z_t^{Y^\beta}\bigr)\psi\bigl(Z_t^{Y^\beta}\bigr)\, dt&=
-\int_0^{t_w}\psi\bigl(Z_t^{Y^\beta}\bigr)\, dZ_t^{Y^\beta}
=-\int_z^{\beta(y)}\psi(u)\, du .
\end{align*}
%Similar calculations verify that
%\begin{gather*}
%\int_0^{t_w}h\bigl(Z_t^{Y^\beta}\bigr)\psi\bigl(Z_t^{Y^\beta}-y\bigr)\, dt
%=-\int_z^{\beta(y)}\psi(u-y)\, du
%=-\int_{z-y}^{\beta(y)-y}\psi(u)\, du .
%\end{gather*}
Hence in case (ii), the performance function is given by 
\begin{align}  \label{JI(iii)calculation}
J_\beta(y,z)&=\int_0^{t_w}\biggl(\kappa_A(y)+A h\bigl(Z_t^{Y^\beta}\bigr)
\psi\bigl(Z_t^{Y^\beta}\bigr)\biggr)\, dt+J_{\beta}\bigl(y,\beta(y)\bigr)\cr
&=\kappa_A(y)\int_{\beta(y)}^z\frac{1}{h(u)}\, du
-A\int_z^{\beta(y)}\psi(u)\, du\cr
&\qquad+\int_{\beta(0+)}^{\gamma_\beta(y)}
\biggl(\frac{\kappa_A\bigl(\gamma_\beta^{-1}(u)\bigr)}{h\bigl(\rho_\beta^{-1}(u)\bigr)}
+A\psi\bigl(\rho_\beta^{-1}(u)\bigr)\biggr)\, du+A\int_0^{\beta(0+)} \psi(u) \,du.
\end{align}
%(($\kappa_A(y)=\infty$ IS EXCLUDED BY Lemma \ref{LemGamma}, SINCE IN CASE (iii) $\beta(y)>-\infty$))
Although this provides an explicit expression for $J_\beta(y,z)$, it is not  entirely straightforward 
to conclude about properties of continuity  and differentiability for $J_\beta(y,z)$ in $y$ 
since $\beta$ is only a \gd function. However, we can further calculate that
\begin{align*} %\label{transcalc1}
\int_{\beta(0+)}^{\gamma_\beta(y)}\biggl(\frac{\kappa_A\bigl(\gamma_\beta^{-1}(u)\bigr)}{h\bigl(\rho_\beta^{-1}(u)\bigr)}
+A\psi\bigl(\rho_\beta^{-1}(u)\bigr)\biggr)\, du
&=\int_{0}^y\biggl(\frac{\kappa_A(u)}{h\bigl(\beta(u)\bigr)}+A\psi\bigl(\beta(u)\bigr)\biggr)\, d\gamma_\beta^c(u)\cr
&\qquad+\sum_{0<u<y} \kappa_A(u)\int_{\beta(u)}^{\beta(u+)}\frac{1}{h(x)}\, dx\cr
&\qquad+A\sum_{0<u< y}\int_{\beta(u)}^{\beta(u+)}\psi(s)\, ds .
\end{align*}
From this expression, as well as
\begin{gather*} %\label{transcalc2}
\int_{\beta(0+)}^{\beta(y)}\psi(u)\, du=\int_{0}^y \psi\bigl(\beta(u)\bigr)\, d\beta^c(u)
+\sum_{0<u<y} \int_{\beta(u)}^{\beta(u+)}\psi(s)\, ds ,
\end{gather*}
and
\begin{align*} %\label{transcalc3}
\kappa_A(y)H\bigl(\beta(y)\bigr)&=\kappa_A(0)H\bigl(\beta(0+)\bigr)+\int_{0}^y \kappa_A'(u) H\bigl(\beta(u)\bigr)\, du
+\int_{0}^y\frac{\kappa_A(u)}{h\bigl(\beta(u)\bigr)}\, d\beta^c(u)\cr
&\qquad+\sum_{0<u<y} \kappa_A(u)\int_{\beta(u)}^{\beta(u+)}\frac{1}{h(x)}\, dx ,
\end{align*}
it follows from (\ref{JI(iii)calculation}) that the performance 
function$J_\beta(y,z)$, for case (ii) of Lemma \ref{Lemmstrat}, 
admits the expression
\begin{align} \label{JI(iii)}
J_\beta(y,z)=\kappa_A(y) H(z)+A\int_{0}^z\psi(u)\, du
-\int_0^y\biggl(\frac{\kappa_A(u)}{h\bigl(\beta(u)\bigr)}+A\psi\bigl(\beta(u)\bigr)
+\kappa_A'(u) H\bigl(\beta(u)\bigr)\biggr) du.
\end{align}
In the above calculations, we have assumed the existence and finiteness 
of $\lim_{u\rightarrow 0+}\frac{\kappa_A(u)}{h(\beta(u))}$ and 
$\lim_{u\rightarrow 0+}\kappa_A'(u) H\bigl(\beta(u)\bigr)$. 
We have also used that $\lim_{u\rightarrow y-}\kappa_A(u)<\infty$ as well as 
$\lim_{u\rightarrow y-}\kappa'_A(u)<\infty$. The finiteness of 
$\lim_{u\rightarrow 0+}\kappa_A'(u) H\bigl(\beta(u)\bigr)$ together with 
(\ref{muneg}) and (\ref{munull}) imply that $\kappa_A(0)H\bigl(\beta(0+)\bigr)=0$. 
For an optimal intervention boundary $\beta$, all of these properties 
will be demonstrated below in Lemma \ref{Lembetaprop}.

%=================================================

Suppose $\beta$ is an intervention boundary such that $Y^\beta$ 
is optimal, i.e. the value function $v$ and the performance function $J_\beta$ coincide. 
Then according to the Hamilton-Jacobi-Bellman 
equation as well as (\ref{JI(iii)}), we should have 
\begin{gather*}
D_y^-v(y,z)+v_z(y,z)=\Gamma(z;y)-\Gamma\bigl(\beta(y);y\bigr)\leq 0 , 
\qquad\text{ for all } (y,z)\in\mathcal{D},
\end{gather*}
%and 
%\begin{gather*}
%D_y^-v(y,z)+v_z(y,z)=0 , \qquad\text{ for } z>\beta(y) ,
%\end{gather*}
where 
\begin{gather*}
\Gamma(x; y)=A\psi(x)+\frac{\kappa_A(y)}{h(x)}+\kappa_A'(y) H(x).
\end{gather*}
Therefore, for any given $y$, $\beta(y)$ should be a 
maximiser of $\Gamma(x; y)$. The next lemma helps us characterise 
an intervention boundary $\beta$ whose value maximises $\Gamma(x; y)$. 
We will later show that such a $\beta$ indeed describes 
an optimal intervention boundary for our problem. 

\begin{lem} \label{LemGamma}
For $y\in (0,\bar{y}_A)$, define the function 
$\Gamma(\cdot;y):[\bar{z},0]\rightarrow \overline{\mathbb{R}}$ by
\begin{gather}\label{Gamma-eq}
\Gamma(x; y)=A\psi(x)+\frac{\kappa_A(y)}{h(x)}+\kappa_A'(y) H(x), 
\qquad\text{ for }x\in(\bar{z},0),
\end{gather}
and 
\begin{gather*}
\Gamma(0;y)=\lim_{x\rightarrow 0+}\Gamma(x;y), 
\qquad \Gamma(\bar{z};y)=\lim_{x\rightarrow \bar{z}}\Gamma(x;y). 
\end{gather*}
%((!!!NEED ASSUMPTION ON $\psi$ and $H$ SO THAT THE SECOND LIMIT EXISTS!!!))
Let $\beta^*=\beta^*(y)$ and $\beta_*=\beta_*(y)$ denote the
functions defined as the largest and smallest $\beta\in [\bar{z},0]$ satisfying
\begin{gather}\label{hequation}
\max_{x\in[\bar{z},0]} \Gamma(x; y)=\Gamma\bigl(\beta; y\bigr), 
\end{gather}
respectively. Then for all $y\in(0,\bar{y}_A)$, we have $\bar{z}\leq\beta_*(y)\leq\beta^*(y)<0$. 
Furthermore, if $\bar{y}_A<\infty$, write $\beta^*(y)=\beta_*(y)=\bar{z}$, for all $y>\bar{y}_A$. Set
\begin{gather*}
\beta^*(0)=0,\qquad\qquad 
\beta_*(0)=\lim_{y\rightarrow 0^+} \beta_*(y) ,
\end{gather*}
and
\begin{gather*}
\beta^*(\bar{y}_A)=\lim_{y\rightarrow \bar{y}_A-} \beta^*(y),\qquad 
\beta_*(\bar{y}_A)=\lim_{y\rightarrow \bar{y}_A+} \beta_*(y) .
\end{gather*}
This uniquely defines two decreasing functions $\beta^*,\beta_*:\mathbb{R}^+\rightarrow [\bar{z},0]$ 
that are \gd and c\`adl\`ag, respectively, and they are the left and the right-continuous version of each other.
\end{lem}

\begin{rem}\label{muposrem}
We will see later that the optimal intervention boundary will be  given by $\beta^*$, and we have previously assumed that $\mu\leq 0$.
But lets examine some of the properties that $\beta^*$ would have if $\mu>0$. If $\mu>0$, then $\kappa_A$ is no longer
strictly increasing, but will be strictly decreasing on $[0,\underline{y})$ and then strictly increasing on $(\underline{y},\bar{y}_A)$,
where $\underline{y}$ denotes the unique point where $\kappa_A$ attains its minimum. In particular $\kappa_A(\underline{y})<0$
and $\kappa_A'(\underline{y})=0$. Therefore
\begin{gather*}
\Gamma(x;\underline{y})=A\psi(x)+\frac{\kappa_A(\underline{y})}{h(x)} ,
\end{gather*}
which is maximised for $x=0$. Thus $\beta^*(\underline{y})=\beta_*(\underline{y})=0$ if $\mu>0$. 
This means that if the optimal intervention boundary is given by $\beta^*$, then the optimal strategy would involve holding on
to at least $\underline{y}$ number of shares until eternity if the initial stock position is larger than $\underline{y}$.
But such a strategy is not admissible as we want to restrict the set of admissible strategies to strategies which tend
to zero sufficiently fast (we would also have to rephrase the admissibility condition if $\mu>0$ since in this case $\kappa_A$
is not strictly increasing). We conclude that if $\mu>0$, then the optimal liquidation problem is not well posed.
\end{rem}

\begin{lem}\label{Gammaconvex}
Let $\beta=\beta^*$ where $\beta^*$ is as in Lemma \ref{LemGamma}.
For $\bar{z}-\bar{y}_A<s<0$ and $\rho_\beta^{-1}(s)<z<0$ it holds that
\begin{gather*}
\Gamma\bigl(\rho_\beta^{-1}(s); \gamma_\beta^{-1}(s)\bigr)-\Gamma\bigl(z; \gamma_\beta^{-1}(s)\bigr)\geq 0 .
\end{gather*}
\end{lem}

Lemma \ref{Gammaconvex} is needed for the proof of Proposition \ref{Propv} (see in 
particular equation (\ref{Gcalc1})). The result relies on the assumptions that
$x\mapsto \nu([x,0])$ and $x\mapsto 1/h(x)$ are concave functions. If we do not
impose these conditions then one can show that the strategy $Y^\beta$ with $\beta=\beta^*$
given by (\ref{hequation}) may not be optimal (compare Theorem \ref{Thmmain}). 
Thus while these concavity conditions may seem realistic, they
are also needed in order to solve the problem. If these conditions does not hold, we simply do
not know what the solution to the liquidation problem looks like.

\begin{lem} \label{Lembetaprop}
Let $\beta^*$ be given by Lemma \ref{LemGamma}, it follows that if 
$\lim_{x\rightarrow y-}\kappa_A(x)=\infty$ or $\lim_{x\rightarrow y-}\kappa_A'(x)=\infty$, then $
\lim_{x\rightarrow y-}\beta^*(x)=\bar{z}$. Furthermore, we have 
\begin{gather}
\lim_{y\rightarrow 0+}\frac{\kappa_A(y)}{h\bigl(\beta^*(y)\bigr)}=0
\qquad\text{ and }\qquad
\lim_{y\rightarrow 0+}\kappa'_A(y)H\bigl(\beta^ *(y)\bigr)=0.  \label{lim-y-to-0}
\end{gather}
\end{lem}
%\noindent
Clearly, the function $\beta^*$ given in Lemma \ref{LemGamma} satisfies 
the properties we require of an intervention boundary. 
With this intervention boundary, the proposition below provides an explicit expression 
for the value function that solves (\ref{h1})--(\ref{h4}) with associated 
boundary condition $v(0,z)=A\int_0^z \psi(u)\,du$, for all $z\in[\bar{z},0]$.
As a consequence, the optimal liquidation strategy is characterised by this intervention boundary. 

\begin{prop}\label{Propv}
%satisfying $\lim_{y\rightarrow \bar{y}_A}\beta(y)=-\infty$.
Let $\beta=\beta^*$ denote the largest solution to (\ref{hequation}),
and let $\gamma_\beta^{-1}$ and $\rho_\beta^{-1}$ be the corresponding
functions defined by (\ref{gammahinv}) and (\ref{rhohinv}). Then the function 
$v:\mathcal{D}\rightarrow\mathbb{R}$ given by  
\begin{gather}\label{vdef1}
v(y,z)=
\int_{\beta(0+)}^{z-y}\biggl(\frac{\kappa_A\bigl(\gamma_\beta^{-1}(u)\bigr)}{h\bigl(\rho_\beta^{-1}(u)\bigr)}
+A\psi\bigl(\rho_\beta^{-1}(u)\bigr)\biggr)\, du +A\int_0^{\beta(0+)} \psi(u) \, du
\end{gather}
for $z>\beta(y)$, and 
\begin{align}\label{vdef2}
v(y,z)&=\kappa_A(y)H(z)
+A\int_{0}^z\psi(u)\, du-\int_0^y\biggl(\frac{\kappa_A(u)}{h\bigl(\beta(u)\bigr)}+A\psi\bigl(\beta(u)\bigr)
+\kappa_A'(u) H\bigl(\beta(u)\bigr)\biggr)\, du ,
\end{align}
for $z\leq\beta(y)$, 
is a $C^{0,1}(\mathcal{D})$ solution to (\ref{h1})--(\ref{h4}) with the boundary condition
\begin{gather*} 
v(0,z)=A\int_0^z\psi(u)\,du,\qquad\text{for all } z\in[\bar{z},0]. 
\end{gather*}
Moreover, $D_y^-v(y,z)$ is 
\gd in $y$ and continuous in $z$. 
\end{prop}

%==================================================================
%\noindent
%Note that if $z-y\geq\beta(0+)$, then for any $u\in[\,\beta(0+),z-y\,]$, 
%by the definitions of $\gamma_\beta^{-1}$ and $\rho_\beta^{-1}$, we have 
%$\gamma_\beta^{-1}(u)=0$ and $\rho_\beta^{-1}(u)=u$. It follows therefore 
%$v(y,z)=0$, when $z-y\geq\beta(0+)$. 
%Note that (\ref{vdef1})-(\ref{vdef2}), with $y=0$, agree with (\ref{JI(i)}). 
The following theorem verifies that the function $v$ given
by (\ref{vdef1})-(\ref{vdef2}) is equal to the value 
function $V$ given by (\ref{Vdef2}), and that the strategy $Y^\beta$ 
corresponding to $\beta=\beta^*$ in Lemma \ref{LemGamma}
is an optimal 
liquidation strategy. Hence, such a $Y^\beta$ provides a solution to the 
utility maximization problem in (\ref{opt}).

\begin{thm}\label{Thmmain}
Let $\beta$ denote the largest solution to (\ref{hequation}), let $v$ be given by (\ref{vdef1}) 
and (\ref{vdef2}), and let $V$ be given by (\ref{Vdef2}). Then $v=V$ on $\mathcal{D}$ and
\begin{gather*}
\sup_{Y\in\mathcal{A}(y)}\mathbb{E}\bigl[U\bigl(C_\infty(Y)\bigr)\bigr]
=-\exp\biggl(-A(c+by)+A\int_{z}^{z-y}\psi(s)\, ds\biggr)\exp\bigl(v(y,z)\bigr) ,
\end{gather*}
where $A$ denotes the agent's risk aversion, $b$ is the initial unaffected 
price, $c$ is the agent's initial cash position, $z=Z_{0-}^Y$ is the initial state of the bid order 
book, and $y$ is the agent's initial share position. The optimal strategy $Y^*$ is equal to $Y^{\beta}\in\mathcal{A}_D(y)$, 
where  $Y^{\beta}$ is the strategy described in Lemma \ref{Lemmstrat} 
corresponding to $\beta$ with $Y_{0-}^{\beta}=y$. 
\end{thm}

\begin{ex}
Suppose that the the bid order book has an equal number $n$ of orders at every price point to a level $\bar{x}$ below the
unaffected best bid price. This corresponds to $x\mapsto m([x,0])=-n(x\wedge\bar{x})$. Therefore $\psi(z)=\frac{1}{n}z$, for
$\bar{z}=n\bar{x}<z\leq 0$, and $-\infty$, for $z\leq \bar{z}$. We calculate that
\begin{gather*}
\Gamma'(x;y)=\frac{A}{n}-\frac{\kappa_A(y)}{\lambda x^2}+\frac{\kappa_A'(y)}{\lambda x} ,\qquad\text{ for }
n\bar{x}<x<0 .
\end{gather*}
In order to find the maximiser, we want to solve $\Gamma'(x;y)=0$. This amounts to solving
\begin{gather}\label{Geq}
\frac{\lambda A}{n}x^2+\kappa_A'(y) x-\kappa_A(y)=0 .
\end{gather}
The unique solution is
\begin{gather*}
x=\frac{n}{2\lambda A}\biggl\{-\kappa_A'(y)-\sqrt{\bigl(\kappa_A'(y)\bigr)^2+\frac{4\lambda A}{n}\kappa_A(y)}\biggr\} .
\end{gather*}
%provided this is greater than $n\bar{x}$.
Since $\Gamma(x;y)$ is concave in $x$ for every $0\leq y<\bar{y}_A$ and equation (\ref{Geq})
has a unique solution for $x\leq 0$, it follows that
\begin{gather}\label{betaex}
\beta(y)=\frac{n}{2\lambda A}\biggl\{-\kappa_A'(y)-\sqrt{\bigl(\kappa_A'(y)\bigr)^2+\frac{4\lambda A}{n}\kappa_A(y)}\biggr\}, 
\qquad\text{for }0\leq y<\bar{y}_A ,
\end{gather}
provided that $\beta(y)>n\bar{x}$.
\begin{itemize}
\item[(i)] ({\bf BM case}) If $L$ is a Brownian motion with drift $\mu\leq 0$, then 
\begin{gather*}
\kappa_A(y)=-A\mu y+\frac{1}{2}A^2\sigma^2y^2,\qquad y\geq 0 ,
\end{gather*}
and hence,
\begin{gather*}
\beta(y)=\frac{n}{2\lambda A}\biggl\{\mu A-\sigma^2A^2y-\sqrt{\mu^2 A^2
-\biggl(\frac{4\lambda\mu A^2}{n}-2\mu\sigma^2A^3\biggr)y
+\biggl(\sigma^4A^4+\frac{2\lambda\sigma^2 A^3}{n}\biggr)y^2}\biggr\} .
\end{gather*}
If $\mu=0$, then it is straightforward to check that the equation for the optimal intervention boundary
coincides with the result obtained in the example in \citet{Lokk}.

\item[(ii)] ({\bf LVG case}) Consider a L\'{e}vy process $L$ with drift $\mu\leq 0$ and a pure jump part with
L\'{e}vy-measure 
\begin{gather}\label{nuex}
\nu(dx)=\begin{cases}
\frac{-1}{\eta\ln(x+1)}(x+1)^{C+D-1}\,dx, \qquad x\in (-1,0),\cr
\frac{1}{\eta\ln(x+1)}(x+1)^{C-D-1}\,dx, \qquad x\in (0,\infty).  
\end{cases}
\end{gather}
where 
\begin{gather}
C=\frac{\theta}{\rho^2}\qquad\text{and}\qquad D=\frac{\sqrt{\theta^2+\frac{2\rho^2}{\eta}}}{\rho^2}.
\end{gather}
If the initial stock price is $1$ (if the initial price is $s\neq 1$, one can deal with this case by
replacing the risk aversion $A$ with $\tilde{A}=sA$) and
\begin{gather}\label{mucond}
\mu=-\frac{1}{\eta}\ln\bigl(1-\frac{\rho^2\eta}{2}-\theta\eta\bigr), 
\end{gather}
then this choice of L\'{e}vy process corresponds to the (linear) L\'{e}vy process approximation of the
exponential variance-gamma process with parameters $(\rho, \eta,\theta)$. 
We refer the reader to \citet{LokkXu} for more details. The common L\'{e}vy
processes considered as models for financial price data assume models of the exponential type, and
taking $L$ to simply be one of these L\'{e}vy processes will in general 
not do a particularly good job as a model for financial price data. We then have
\begin{align*}
\kappa_A(y)&=-A\mu y+\int_{(-1,\infty)\setminus\{0\}}\biggl(\mathrm{e}^{-Ayz}-1+Ayz\biggr)\,\nu(dz) ,\cr
\kappa_A'(y)&=-A\mu+A\int_{(-1,\infty)\setminus\{0\}}\biggl(1-\mathrm{e}^{-Ayz}\biggr)z\,\nu(dz) ,
\end{align*}
where $\nu$ is given by (\ref{nuex}). With reference to (\ref{betaex}) we can calculate the optimal
intervention boundary $\beta$.
\end{itemize}

\begin{figure}
\begin{center}
\includegraphics[scale=0.6]{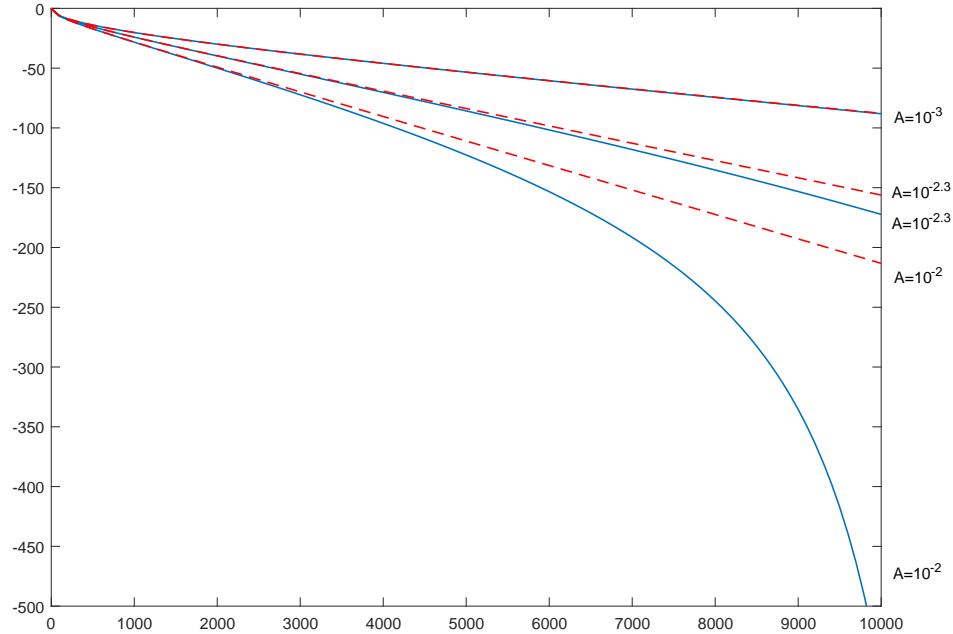}
\end{center}
\caption{The optimal intervention boundary $\beta$ for the two cases of $L$ being a Brownian motion (BM case)
and $L$ being the L\'{e}vy process approximating the exponential variance-gamma (LVG case). 
We take $n=1000$ and $\lambda=5$.
For the LVG case, we take
$\theta=-0.002$, $\rho=0.02$ and $\eta=0.6$ (timescale in days),
which according to \citet{RSS} are realistic parameters for stock price data for the exponential
variance-gamma model. We choose $\mu=-0.0018$, which is in accordance with (\ref{mucond}).
For the BM case, we also choose $\mu=-0.0018$, and take $\sigma^2=4.011*10^{-4}$, which match the
variance of $L$ in the LVG case. Solid graphs correspond to the LVG case and a dotted graphs to the BM case.
When $A=10^{-3}$ we see that the LVG and the BM graphs are pretty much identical.
%The solid graph at the bottom is $\beta$
%for the LVG case with $A=10^{-3}$ and the dotted line is the graph of $\beta$ for the BM case with $A=10^{-3}$.
%The top graph is the graph of both the LVG and the BM case for $A=10^{-4}$ (the graphs pretty much coincide
%in this picture). 
}\label{exfig}
\end{figure}
\noindent
From Figure \ref{exfig} we see that the difference between the BM case and the LVG case increase with the
agent's share position. Moreover the difference becomes more pronounced the more risk averse the agent is.
If the agent is not particularly risk averse, then the more heavy-tailed risk in the LVG model does not make
much difference to the agent, but if the agent is risk averse then it does. However, from Figure \ref{exfig} we see
that while for $A=10^{-2}$ the optimal intervention boundary for the LVG case and the BM case are quite different,
in terms of the corresponding strategy $Y^\beta$, the initial block sale in the LVG case is about $200$ shares and
in the BM case about $150$ shares, if the initial stock position is $10^{4}$. An initial block sale of $200$ shares
would eat into the order book and reduce the best bid price by $\frac{200}{1000}$, so the best bid immediately
after the block sale would be $0.8$ (remember that we assumed the initial stock price to be $1$). 
In the case of an initial block sale of $150$ shares the best bid would be $1-\frac{150}{1000}=0.85$ immediately
after the block sale. 
\end{ex}

\section{Proofs}

\begin{proof}[\textbf{Proof of Lemma \ref{LemmC}}]
With reference to the dynamic of $Z^Y$, we calculate that for $z\geq\bar{z}$, 
\begin{align*}
\int_0^{Z^Y_T} \psi(u)\,du&=\int_0^z \psi(u)\,du+\int_0^T \psi(Z^Y_{t-})\,dY^c_t\cr
&\qquad-\int_0^T h(Z^Y_{t-})\psi(Z^Y_{t-})\,dt
+\sum_{0\leq t\leq T}\int_{Z^Y_{t-}}^{Z^Y_{t-}+\triangle Y_t} \psi(u)\,du\cr
&=\int_0^z \psi(u)\,du+\int_0^T \psi(Z^Y_{t-})\,dY^c_t\cr
&\qquad-\int_0^T h(Z^Y_{t-})\psi(Z^Y_{t-})\,dt
+\sum_{0\leq t\leq T}\int_{0}^{\triangle Y_t} \psi\bigl(Z^Y_{t-}+u\bigr)\,du. 
\end{align*}
Then, 
\begin{align*}
F_T(Y)&=\int_0^T \psi\bigl(Z_{t-}^Y\bigr)\, dY_t^c
+\sum_{0\leq t\leq T}\int_{0}^{\triangle Y_t}\psi\bigl(Z_{t-}^Y+x\bigr)\, dx\cr
&=\int_0^{Z^Y_T} \psi(u)\,du-\int_0^z \psi(u)\,du+\int_0^T h(Z^Y_{t-})\psi(Z^Y_{t-})\,dt\cr
&=\int_z^{Z^Y_T} \psi(u)\,du+\int_0^T h(Z^Y_{t-})\psi(Z^Y_{t-})\,dt. 
\end{align*}
Notice that for any admissible liquidation strategy $Y$, we have that either $Y$ and $Z^Y$ become $0$ 
at the same time or $Y$ becomes $0$ at some time $s$ while $Z^Y_s<0$. In the second case, for all $t>s$, 
$Z^Y$ satisfies 
\begin{gather*}
dZ^Y_t=-h(Z^Y_t)\,dt. 
\end{gather*}
According to (\ref{Zt-in-terms-of-H}), we know that the solution to the above dynamic tends to 0, 
as $t\rightarrow\infty$. Therefore, $Z^Y_t\rightarrow 0$, as $t\rightarrow\infty$ in any case. 
It then follows from the above expression for $F_T(Y)$ that 
\begin{gather*}
F_\infty(Y)=\int_z^{0} \psi(u)\,du+\int_0^\infty h(Z^Y_{t-})\psi(Z^Y_{t-})\,dt. 
\end{gather*}
\end{proof}

%-----------------------------------------------------------------

\begin{proof}[\textbf{Proof of Lemma \ref{Lemmstrat}}]
We first prove that on any time interval $I$ contained in 
$[\,t_w, \bar{t}\,]\setminus\,\cup_{ n\in \mathbb{I}}[t_n,s_n)$, 
there exists a unique solution to the dynamic (\ref{dynamic-of-Y}). 
On such an interval $I$, the process $Y^\beta$ does not cross any jump of $\beta$.  
Thus, in terms of the function $\beta$, we shall only focus on those parts without jumps. Also, it is sufficient to consider $Y$ starting from time 0 (rather than starting at any time in $[\,t_w, \bar{t}\,]\setminus\,\cup_{ n\in \mathbb{I}}[t_n,s_n)$). 
Write $Y_t^0=Y_0>0$ and 
\begin{gather}
Y_t^{k+1}=\gamma_\beta^{-1}\biggl(\Bigl\{\gamma_\beta(Y_0)
-\int_0^t h\bigl(\beta(Y_u^k)\bigr)\,du\Bigr\}\,\wedge\,\beta(0+)\biggr). \label{Yiteration}
\end{gather}
Let $T\in[0,\infty)$. Then 
\begin{align}  \label{iteration}
&\sup_{0\leq t\leq T}\bigl|\,\beta(Y_t^{k+1})-\beta(Y_t^{k})\,\bigr|\cr
=&\sup_{0\leq t\leq T}\biggl|\Bigl\{\gamma_\beta(Y_0)
  -\int_0^t h\bigl(\beta(Y_u^k)\bigr)\,du\Bigr\}\wedge\beta(0+)
  -\Bigl\{\gamma_\beta(Y_0)
  -\int_0^t h\bigl(\beta(Y_u^{k-1})\bigr)\,du\Bigr\}\wedge\beta(0+)\cr
  &\qquad\qquad+\gamma_\beta^{-1}\biggl(\Bigl\{\gamma_\beta(Y_0)
    -\int_0^t h\bigl(\beta(Y_u^k)\bigr)\,du\Bigr\}\wedge\beta(0+)\biggr)\cr
  &\qquad\qquad-\gamma_\beta^{-1}\biggl(\Bigl\{\gamma_\beta(Y_0)
  -\int_0^t h\bigl(\beta(Y_u^{k-1})\bigr)\,du\Bigr\}\wedge\beta(0+)\biggr)\biggr|\cr
\leq&\,2\sup_{0\leq t\leq T}\biggl|\,\Bigl\{\gamma_\beta(Y_0)
  -\int_0^t h\bigl(\beta(Y_u^k)\bigr)\,du\Bigr\}\wedge\beta(0+)
  -\Bigl\{\gamma_\beta(Y_0)
  -\int_0^t h\bigl(\beta(Y_u^{k-1})\bigr)\,du\Bigr\}\wedge\beta(0+)\,\biggr|\cr
\leq&\,2\,\sup_{0\leq t\leq T}\biggl|\,\int_0^t h\bigl(\beta(Y_u^k)\bigr)-h\bigl(\beta(Y_u^{k-1})\bigr)\,du\,\biggr|\cr
\leq&\,2L\,\int_0^T\Bigl|\, \beta(Y_u^k)-\beta(Y_u^{k-1})\,\Bigr|\,du\cr
\leq&\,2L\,\int_0^T\sup_{0\leq t\leq u}\Bigl|\, \beta(Y_t^k)-\beta(Y_t^{k-1})\,\Bigr|\,du,
\end{align}
where the first equality holds because when $\beta$ has no 
jumps we have $\beta\bigl(\gamma_\beta^{-1}(x)\bigr)=x+\gamma_\beta^{-1}(x)$, 
the first inequality is due to the triangle inequality and 
$|\gamma_\beta^{-1}(x)-\gamma_\beta^{-1}(y)|\leq |x-y|$, 
and the third inequality follows from the boundedness of the
processes $\beta(Y^k)$ and $\beta(Y^{k-1})$ and the local 
Lipschitz continuity of $h$ with a Lipschitz constant $L$. By 
induction and with reference to (\ref{iteration}), it can be 
shown that 
\begin{gather*}
\sup_{0\leq t\leq T}\bigl|\,\beta(Y_t^{k+1})-\beta(Y_t^{k})\,\bigr|
\leq \frac{(2LT)^{k}}{k!}2\bigl|\beta(Y_0)\bigr|.
\end{gather*}
Taking $k$ to infinity, we have that $\beta(Y_t^k)$ converges 
uniformly on $[0,T]$. Define $\beta_t=\lim_{k\rightarrow\infty}\beta(Y_t^k)$, 
for $t\in[0,T]$. Since $T\in[0,\infty)$ is arbitrary, it follows that 
$\beta_t=\lim_{k\rightarrow\infty}\beta(Y_t^k)$ for all $t\in[0,\infty)$. 
With reference to (\ref{Yiteration}) and the dominated convergence theorem it follows that,
for every $t\in[0,\infty)$, $\bigl(Y_t^k\bigr)_{k=0}^\infty$ is convergent. 
We define $Y_t^\beta=\lim_{k\rightarrow\infty}Y_t^k$. It can be checked that $Y^\beta$ 
decreases to 0. Then since $\beta$ is continuous, we obtain $\beta_t=\beta(Y_t^\beta)$, 
for all $t\in[0,\infty)$. Therefore, by sending $k$ to infinity in (\ref{Yiteration}), 
since we only consider $Y_t^\beta$ before time $\bar{t}$, we have that 
\begin{gather*}
Y^\beta_t=\gamma_\beta^{-1}\biggl(\gamma_\beta(Y^\beta_0)-\int_0^t h\bigl(\beta(Y^\beta_u)\bigr)\, du\biggr)
,\quad \text{ for }t\leq \bar{t} .
\end{gather*}
This proves the existence of a solution to the dynamic (\ref{dynamic-of-Y}) 
on any time interval contained in $[\,t_w, \bar{t}\,]\setminus\,\cup_{ n\in \mathbb{I}}[t_n,s_n)$. 
For uniqueness, let's assume that $Y^{(1)}$ and $Y^{(2)}$ satisfy 
(\ref{dynamic-of-Y}), where $Y_t^{(1)}=Y_t^{(2)}$ for $0\leq t\leq t_1$, 
and $Y_t^{(1)}<Y_t^{(2)}$ for $t_1<t<t_2$. Then for $t_1<t<t_2$,
\begin{align*}
Y_t^{(1)}&=\gamma_\beta^{-1}\biggl(\gamma_\beta(Y_0^{(1)})-\int_0^t h\bigl(\beta(Y_u^{(1)})
\bigr)\, du\biggr)\cr
&\geq\gamma_\beta^{-1}\biggl(\gamma_\beta(Y_0^{(2)})-\int_0^t h\bigl(\beta(Y_u^{(2)})
\bigr)\, du\biggr)\cr
&=Y_t^{(2)} ,
\end{align*}
which contradicts the assumption that $Y_t^{(1)}<Y_t^{(2)}$
for $t_1<t<t_2$. So uniqueness holds. The existence and 
uniqueness of a solution to the dynamic in (\ref{dynamic-of-Z}) 
on any time interval contained in $[\,t_w, \bar{t}\,]\cap\,\cup_{ n\in \mathbb{I}}[t_n,s_n)$
follow from the 
%Picard-Lindel\"{o}f theorem \citep[see e.g.][]{CL} 
%provided with assumption that $h$ is 
locally Lipschitz continuity of the function $h$. 

Now let $Y^\beta$ and $Z^{Y^\beta}$ be processes satisfying 
(\ref{Y=beta^-1(Z)})--(\ref{Ztbar-initial}) with 
$\bigl(Y_{0-}^\beta, Z_{0-}^{Y^\beta}\bigr)=(y,z)\in\mathcal{D}$. Note that 
$\bigl(Y_t^\beta, Z_t^{Y^\beta}\bigr)\in\mathcal{G}^\beta$ for all $t\in [\,t_w,\bar{t}\,]$. 
We need to show that (\ref{Zeq}) is satisfied. We first focus on the case when 
$t\leq t_w$. Suppose $z>\beta(y)$, i.e. $t_w=0$. Then in case (i) (a), 
\begin{align*}
Y_{0}^\beta-Y_{0-}^\beta&=\beta^{-1}\bigl(\rho^{-1}_\beta(z-y)\bigr)-y\cr
&=\gamma^{-1}_\beta(z-y)-y\cr
&=\bigl(z-y+\gamma^{-1}_\beta(z-y)\bigr)-z\cr
&=Z_{0}^{Y^\beta}-Z_{0-}^{Y^\beta}, 
\end{align*}
where we have used the identity $\beta^{-1}\bigl(\rho_\beta^{-1}(z-y)\bigr)=\gamma_\beta^{-1}(z-y)$, 
which follows from (\ref{iden2}) and is valid under the condition of (i) (a). 
In case (i) (b), we obtain 
\begin{align*}
Z_{0}^{Y^\beta}-Z_{0-}^{Y^\beta}
&=\beta\bigl(\gamma^{-1}_\beta(z-y)\bigr)-z\cr
&=z-y+\gamma^{-1}_\beta(z-y)-z\cr
&=\gamma^{-1}_\beta(z-y)-y\cr
&=Y_{0}^\beta-Y_{0-}^\beta, 
\end{align*}
where we have used that $\beta\bigl(\gamma_\beta^{-1}(z-y)\bigr)=\rho_\beta^{-1}(z-y)$. 
Suppose $z\leq\beta(y)$, i.e. $t_w>0$. It can be checked that 
$Z_t^{Y^\beta}=H^{-1}\bigl(H(z)-t\bigr)$ has dynamic (\ref{dynamic-of-Z}). 
Because $Y^\beta_t$ is now constant, (\ref{Zeq}) is satisfied. 
In the case when $t>t_w$, $Y^\beta_t$ and $Z_t^{Y^\beta}$ follow 
(\ref{Y=beta^-1(Z)})--(\ref{Ztbar-initial}), which satisfy (\ref{Zeq}).

We next like to prove that $Y^\beta$ is \dg and decreasing. Note that 
by the definitions of $t_n$, $s_n$, $t_w$ and $\bar{t}$ and 
(\ref{dynamic-of-Z}), (\ref{dynamic-of-Y}) and the first part 
of the proof, we have $Y_t^\beta$ and $Z_t^{Y^\beta}$ are 
continuous when $(Y_t^\beta, Z_t^{Y^\beta})$ is in each continuous 
part of the graph of $\beta$ or $\beta^{-1}$, for $t>0$. 
Also, each initial condition associated with the dynamics (\ref{dynamic-of-Z}) 
and (\ref{dynamic-of-Y}) is chosen to make $Y_t^\beta$ and 
$Z_t^{Y^\beta}$ to be continuous at $t_n$, $s_n$ and $t_w$ when 
$t_w>0$. It can also be seen that $Y^\beta$ and $Z^{Y^\beta}$ are 
right continuous at $t=0$. These together with the well-defined $Y^\beta_{0-}$ 
and $Z^{Y^\beta}_{0-}$ imply that $Y^\beta$ and $Z^{Y^\beta}$ 
are continuous for $t>0$ and they are right-continuous with left-limit 
at $t=0$. That $Y^\beta$ decreases to 0 follows from (\ref{Y=beta^-1(Z)}), 
(\ref{dynamic-of-Z}), (\ref{dynamic-of-Y}), and the first part of 
this proof. 
Finally, that $Z_t^{Y^\beta}=H^{-1}\bigl(H(z)-t\bigr)$, for $0\leq t\leq t_w$, 
follows from (\ref{Zt-in-terms-of-H}). 
\end{proof}

%-------------------------------------------------------------------------

\begin{proof}[\textbf{Proof of Lemma \ref{Lemmtw-infty-rewite}}]
Let $\{y_n\}_{n\in\mathbb{I}}$ be the set of all points at which 
the intervention boundary $\beta$ is discontinuous. 
Consider a time interval $[t,s]\subseteq[t_n,s_n)$, 
for some $n\in\mathbb{I}$, where $t_n$ and $s_n$ are given by (\ref{tn}) and 
(\ref{sn}). With reference to (\ref{Zeq}), we note that formally,
\begin{gather*}
dt=-\frac{d\rho_\beta\bigl(Z_t^{Y^\beta}\bigr)}{h\bigl(Z_{t}^{Y^\beta}\bigr)}\qquad\forall\,t\in[t_n,s_n),
\end{gather*}
and hence, 
\begin{align}  \label{s-t-in-[tn,sn)}
&\quad\,\int_t^s \biggl(\kappa_A\bigl(Y_{r}^\beta\bigr)+A h\bigl(Z_{r}^{Y^\beta}\bigr)
\psi\bigl(Z_{r}^{Y^\beta}\bigr)\biggr)\, dr\cr
&=\int_s^t \biggl(\frac{\kappa_A\bigl(\beta^{-1}(Z_{r}^{Y^\beta})\bigr)}{h(Z_{r}^{Y^\beta})}
+A\psi\bigl(Z_{r}^{Y^\beta}\bigr)
\biggr)\, d\rho_\beta\bigl(Z_{r}^{Y^\beta}\bigr) \cr
&=\int_{\rho_\beta(Z_{s}^{Y^\beta})}^{\rho_\beta(Z_{t}^{Y^\beta})}
\biggl(\frac{\kappa_A\bigl(\gamma_\beta^{-1}(u)\bigr)}{h\bigl(\rho_\beta^{-1}(u)\bigr)}
+A\psi\bigl(\rho_\beta^{-1}(u)\bigr)\biggr)\, du \cr
&=\int_{Z_{s}^{Y^\beta}-Y^\beta_{s_n}}^{Z_{t}^{Y^\beta}-Y^\beta_{t_n}}
\biggl(\frac{\kappa_A\bigl(\gamma_\beta^{-1}(u)\bigr)}{h\bigl(\rho_\beta^{-1}(u)\bigr)}
+A\psi\bigl(\rho_\beta^{-1}(u)\bigr)\biggr)\, du ,
\end{align}
where we have used the identity in (\ref{iden2}). 
%((It is worth mentioning here that for all $n\in\mathbb{I}$, 
%$Y^\beta_{s_n}=Y^\beta_{t_n}$. This is because of the definitions 
%of $s_n$ and $t_n$ and the fact that $Y^\beta_t$ is continuous when 
%$t\geq 0$ (see the proof of Lemma \ref{Lemmstrat}).)) 
Similarly, since
\begin{gather*}
dt=-\frac{d\gamma_\beta\bigl(Y_t^\beta\bigr)}{h\bigl(\beta({Y_t^\beta})\bigr)}\qquad
\forall\,t\in[\,t_w, \bar{t}\,]\setminus\,\cup_{ n\in \mathbb{I}}[t_n,s_n),
\end{gather*}
applying (\ref{iden3}), it can be calculated that on some time interval 
$[s,t]\subset[\,t_w, \bar{t}\,]\setminus\,\cup_{ n\in \mathbb{I}}[t_n,s_n)$, for some $n\in\mathbb{I}$, 
\begin{align}  \label{s-t-in-non-[tn,sn)}
&\quad\,\int_s^t \biggl(\kappa_A\bigl(Y_{r}^\beta\bigr)+A h\bigl(Z_{r}^{Y^\beta}\bigr)
\psi\bigl(Z_{r}^{Y^\beta}\bigr)\biggr)\, dr\cr
%&=\int_t^s \biggl(\frac{\kappa_A\bigl(Y_{r}^\beta\bigr)}{h\bigl(\beta(Y_r^\beta)\bigr)}
%+A\psi\bigl(\beta(Y_r^\beta)\bigr)
%-A\psi\bigl(\gamma_\beta(Y_r^\beta)\bigr)\biggr)\, d\gamma_\beta\bigl(Y_r^\beta\bigr) \cr
%&=\int_{\gamma_\beta(Y_t^\beta)}^{\gamma_\beta(Y_s^\beta)}
%\biggl(\frac{\kappa_A\bigl(\gamma_\beta^{-1}(u)\bigr)}{h\bigl(\rho_\beta^{-1}(u)\bigr)}
%+A\psi\bigl(u+\gamma_\beta^{-1}(u)\bigr)-A\psi(u)\biggr)\, du \cr
&=\int_{Z_{t}^{Y^\beta}-Y^\beta_{t}}^{Z_{s}^{Y^\beta}-Y^\beta_{s}}
\biggl(\frac{\kappa_A\bigl(\gamma_\beta^{-1}(u)\bigr)}{h\bigl(\rho_\beta^{-1}(u)\bigr)}
+A\psi\bigl(\rho_\beta^{-1}(u)\bigr)\biggr)\, du .
\end{align}
Let $t_w<\bar{t}$. Suppose the number of $t_n$ and $s_n$ in the interval 
$[\,t_w, \bar{t}\,]$ is equal to $m<\infty$ (possibly $m=0$). Consider $r_0\leq r_1<...<r_m<r_{m+1}$, 
where $r_0=t_w$, $r_{m+1}=\bar{t}$ and for $k=1,...,m$, $r_k$ are equal to those 
$t_n,s_n\in[\,t_w, \bar{t}\,]$. We assume $r_1,...,r_m$ are in an ascending order. 
Then it follows from (\ref{s-t-in-[tn,sn)}), (\ref{s-t-in-non-[tn,sn)}) and the 
continuity of $Y^\beta_t$ and $Z^{Y^\beta}_t$ when $t>0$ that 
\begin{align*}
&\quad\,\int_{t_w}^{\bar{t}} \biggl(\kappa_A\bigl(Y_{t}^\beta\bigr)+A h\bigl(Z_{t}^{Y^\beta}\bigr)
\psi\bigl(Z_{t}^{Y^\beta}\bigr)\biggr)\, dt\cr
&=\sum_{k=0}^m \int_{r_{k}}^{r_{k+1}} 
\biggl(\kappa_A\bigl(Y_{t}^\beta\bigr)+A h\bigl(Z_{t}^{Y^\beta}\bigr)
\psi\bigl(Z_{t}^{Y^\beta}\bigr)\biggr)\, dt\cr
&=\int_{Z_{\bar{t}}^{Y^\beta}}^{Z_{t_w}^{Y^\beta}-Y^\beta_{t_w}}
\biggl(\frac{\kappa_A\bigl(\gamma_\beta^{-1}(u)\bigr)}{h\bigl(\rho_\beta^{-1}(u)\bigr)}
+A\psi\bigl(\rho_\beta^{-1}(u)\bigr)\biggr)\, du.
\end{align*}
Suppose there are infinitely many $t_n$ and $s_n$ in the interval $[\,t_w, \bar{t}\,]$. 
Let $r\in[\,t_w,\bar{t}\,]$ be an accumulation point of the sequence $\{t_n\}_{n\in\mathbb{I}}$. 
Then without loss of generality, consider a subsequence $\{t_{n_k}\}_{k=1}^\infty\subset[\,t_w,\bar{t}\,]$ 
increasing to $r$. Consider some time interval $[t,s]$ in which $r$ is the only accumulation point of 
$\{t_n\}_{n\in\mathbb{I}}$. Then, it follows that
\begin{align*}
&\quad\,\int_{t}^{s} \biggl(\kappa_A\bigl(Y_{t}^\beta\bigr)+A h\bigl(Z_{t}^{Y^\beta}\bigr)
\psi\bigl(Z_{t}^{Y^\beta}\bigr)\biggr)\, dt\cr
&=\lim_{n\rightarrow\infty}\int_{t}^{t_n} \biggl(\kappa_A\bigl(Y_{t}^\beta\bigr)
+A h\bigl(Z_{t}^{Y^\beta}\bigr)\psi\bigl(Z_{t}^{Y^\beta}\bigr)\biggr)\, dt+\int_{r}^{s} \biggl(\kappa_A\bigl(Y_{t}^\beta\bigr)
+A h\bigl(Z_{t}^{Y^\beta}\bigr)\psi\bigl(Z_{t}^{Y^\beta}\bigr)\biggr)\, dt\cr
&=\lim_{n\rightarrow\infty}\int_{Z_{t_n}^{Y^\beta}-Y^\beta_{t_n}}^{Z_{t}^{Y^\beta}-Y^\beta_{t}}
\biggl(\frac{\kappa_A\bigl(\gamma_\beta^{-1}(u)\bigr)}{h\bigl(\rho_\beta^{-1}(u)\bigr)}
+A\psi\bigl(\rho_\beta^{-1}(u)\bigr)\biggr)\, du\cr
&\qquad+\int_{Z_{s}^{Y^\beta}-Y^\beta_{s}}^{Z_{r}^{Y^\beta}-Y^\beta_{r}}
\biggl(\frac{\kappa_A\bigl(\gamma_\beta^{-1}(u)\bigr)}{h\bigl(\rho_\beta^{-1}(u)\bigr)}
+A\psi\bigl(\rho_\beta^{-1}(u)\bigr)\biggr)\, du,\cr
&=\int_{Z_{s}^{Y^\beta}-Y^\beta_{s}}^{Z_{t}^{Y^\beta}-Y^\beta_{t}}
\biggl(\frac{\kappa_A\bigl(\gamma_\beta^{-1}(u)\bigr)}{h\bigl(\rho_\beta^{-1}(u)\bigr)}
+A\psi\bigl(\rho_\beta^{-1}(u)\bigr)\biggr)\, du.
\end{align*}
%((Note that because $\beta$ is \gd, there are at most countably many accumulative points 
%of $\{t_n\}_{n\in\mathbb{I}}$. COUNTABLY MANY IS FOR THE PROOF OF ADMISSIBILITY)) 
This implies that 
\begin{align*}
&\quad\,\int_{t_w}^{\bar{t}} \biggl(\kappa_A\bigl(Y_{t}^\beta\bigr)+A h\bigl(Z_{t}^{Y^\beta}\bigr)
\psi\bigl(Z_{t}^{Y^\beta}\bigr)\biggr)\, dt\cr
&=\int_{Z_{\bar{t}}^{Y^\beta}}^{Z_{t_w}^{Y^\beta}-Y^\beta_{t_w}}
\biggl(\frac{\kappa_A\bigl(\gamma_\beta^{-1}(u)\bigr)}{h\bigl(\rho_\beta^{-1}(u)\bigr)}
+A\psi\bigl(\rho_\beta^{-1}(u)\bigr)\biggr)\, du.
\end{align*}
Therefore the result follows from the above equality as well as (\ref{Ztbar-initial}) and (\ref{tbar-infty=0}).
\end{proof}

%----------------------------------------------------------------------

\begin{proof}[\textbf{Proof of Lemma \ref{LemGamma}}]
%By (\ref{phipsicdt3})and (\ref{Hdef}), we have 
%\begin{gather*}
%\lim_{x\rightarrow-\infty}\frac{\psi(x)}{x}>0=-C\lim_{x\rightarrow-\infty}\frac{H(x)}{x}, 
%\end{gather*}
%where $C>0$ is some constant. Then, 
First notice that, for any $y\in(0,\bar{y}_A)$, the function $\Gamma(x,y)$ is concave in $x$, but that this concavity may 
not be strict. 
%, since each of the functions $\psi$, $\frac{1}{h}$ and $H$ may not be strictly concave; 
%and in particular, the concavity of $\frac{\kappa_A(y)}{h(x)}+\kappa'_A(y)H(x)$ is not strict, if $h'(x)=0$. 
%Observe that (\ref{phipsicdt3}) gives out 
%\begin{gather*}
%\lim_{x\rightarrow-\infty}\frac{\psi(x)+CH(x)}{x}>0,
%\end{gather*}
%where $C>0$ is some constant. This implies that 
%\begin{gather*}
%\lim_{x\rightarrow-\infty}\psi(x)+CH(x)=-\infty.
%\end{gather*}
Observe that for $y\in(0,\bar{y}_A)$, 
\begin{gather*}
\lim_{x\rightarrow 0^-}\Gamma(x;y)=-\infty .
\end{gather*}
Also, $\Gamma(x;y)\in\mathbb{R}$, for $x\in[\bar{z},0)$. 
%\begin{gather}  \label{Gamma(-infty,y)}
%\lim_{x\rightarrow\bar{z}}\Gamma(x;y)=-\infty. 
%%\leq A\lim_{x\rightarrow -\infty}\biggl(\psi(x)+\frac{\kappa'_A(y)}{A}H(x)\biggr)
%\end{gather}
%Since $\psi$ is right-continuous and increasing, 
%then for $y\in (0,\bar{y}_A)$, $\Gamma(\cdot;y)$ is upper semi-continuous. 
These observations imply that $\bar{z}\leq\beta_*(y)\leq\beta^*(y)<0$, for all $ 0<y<\bar{y}_A$. 
The largest and smallest solution to (\ref{hequation}) uniquely define the functions $\beta^*$ 
and $\beta_*$. For $0<y<y+\triangle<\bar{y}_A$ and $x\in[\bar{z},0)$, 
we calculate that
%\begin{gather}\label{Gamma-difference}
%\Gamma(x;y+\triangle)-\Gamma(x;y)
%=\int_y^{y+\triangle} \biggl(\frac{\kappa_A'(u)}{h(x)}+\kappa_A''(u)H(x)\biggr)\, du ,
%\end{gather}
%and
\begin{gather}\label{Gamma-der}
\frac{d}{dx}\biggl[\Gamma(x;y+\triangle)-\Gamma(x;y)\biggr]
=-\frac{\bigl(\kappa_A(y+\triangle)-\kappa_A(y)\bigr)h'(x)}{h^2(x)}
+\frac{\kappa'_A(y+\triangle)-\kappa'_A(y)}{h(x)}<0, 
\end{gather}
%((NEED h IS ABSOLUTELY CONTINUOUS W.R.T. x)) 
since $\kappa_A$ is convex and $\kappa_A'(u)>0$, for $u>0$.
We want to show that $\beta^*$ and $\beta_*$ are decreasing 
functions. In order to get a contradiction, suppose that 
there exists $y\in(0,\bar{y}_A)$ and $\triangle>0$ such that
$\beta^*(y+\triangle)> \beta_*(y)$. With reference to (\ref{Gamma-der}), 
we obtain 
\begin{align*}
\Gamma\bigl(\beta^*(y+\triangle); y+\triangle\bigr)-\Gamma\bigl(\beta^*(y+\triangle); y\bigr)
<\Gamma\bigl(\beta_*(y); y+\triangle\bigr)-\Gamma\bigl(\beta_*(y); y\bigr) .
\end{align*}
However, this contradicts the definitions of $\beta^*$ and $\beta_*$, which imply that
\begin{align*}
\Gamma\bigl(\beta^*(y+\triangle); y+\triangle\bigr)\geq \Gamma\bigl(\beta_*(y); y+\triangle\bigr)
\quad \text{ and }\quad
\Gamma\bigl(\beta_*(y); y\bigr)\geq \Gamma\bigl(\beta^*(y+\triangle); y\bigr) .
\end{align*}
Therefore, for all $0<y<\bar{y}_A$, 
\begin{gather}\label{betaineq}
\beta_*(y+\triangle)\leq \beta^*(y+\triangle)\leq\beta_*(y)\leq\beta^*(y) ,
\end{gather}
from which it follows that $\beta^*$ and $\beta_*$ are decreasing. 
By (\ref{Gamma-eq}), we know that for $\bar{z}\leq x<0$, 
$\Gamma(x;y)$ is continuous in $y$. Then for $y\in(0,\bar{y}_A)$, 
we have 
\begin{align*}
\Gamma\bigl(\beta_*(y+);y+\bigr)=\Gamma\bigl(\beta_*(y+);y\bigr) \leq\Gamma\bigl(\beta_*(y);y\bigr)=\Gamma\bigl(\beta_*(y);y+\bigr)\cr
\Gamma\bigl(\beta^*(y-);y\bigr)=\Gamma\bigl(\beta^*(y-);y-\bigr)
\geq \Gamma\bigl(\beta^*(y);y-\bigr)=\Gamma\bigl(\beta^*(y);y\bigr).
\end{align*}
Since $\beta^*$ and $\beta_*$ are defined as respectively the largest and 
smallest maximiser to (\ref{hequation}), and
$\beta^*$ and $\beta_*$ are decreasing, it follows 
that $\beta_*(y+)=\beta_*(y)$ and $\beta^*(y-)=\beta^*(y)$. By 
monotonicity, the right limit of $\beta^*$ and the left limit of $\beta_*$ 
exist. Hence, we have proved that $\beta^*$ is \gd and $\beta_*$ 
is \dg. The claim that 
$\beta^*$ is the \gd version of $\beta_*$ and that $\beta_*$ is 
the \dg version of $\beta^*$ follows from (\ref{betaineq}). 
\end{proof}

\begin{proof}[\textbf{Proof of Lemma \ref{Gammaconvex}}]

With reference to (\ref{iden3}), we have that
\begin{gather*}
\beta\bigl(\gamma_\beta^{-1}(s)\bigr)=\rho_\beta^{-1}\bigl(\gamma_\beta\bigl(\gamma_\beta^{-1}(s)\bigr)\bigr)=\rho_\beta^{-1}(s)
\qquad\text{ if }\gamma_\beta\bigl(\gamma_\beta^{-1}(s)\bigr)=s .
\end{gather*}
Moreover, $\gamma_\beta\bigl(\gamma_\beta^{-1}(s)\bigr)=s$, unless $\beta$ has a jump at $\gamma_\beta^{-1}(s)$.
Thus if $\beta$ does not have a jump at $\gamma_\beta^{-1}(s)$ then
\begin{gather}
\Gamma\bigl(\rho_\beta^{-1}(s);\gamma_\beta^{-1}(s)\bigr)
-\Gamma\bigl(z;\gamma_\beta^{-1}(s)\bigr)
=
\Gamma\bigl(\beta\bigl(\gamma_\beta^{-1}(s)\bigr);\gamma_\beta^{-1}(s)\bigr)
-\Gamma\bigl(z;\gamma_\beta^{-1}(s)\bigr)
\geq 0 ,
\end{gather}
by the definition of $\beta$. If on the other hand $\beta$ has a jump at $\gamma_\beta^{-1}(s)$, then
$\gamma_\beta^{-1}$ is flat on the interval $[s_*,s^*]$, where
$s_*=\beta\bigl(\gamma_\beta^{-1}(s)+\bigr)-\gamma_\beta^{-1}(s)$ and
$s^*=\beta\bigl(\gamma_\beta^{-1}(s)\bigr)-\gamma_\beta^{-1}(s)$.
Also
\begin{gather*}
\beta\bigl(\gamma_\beta^{-1}(s_*)+\bigr)=\rho_\beta^{-1}\bigl(\gamma_\beta\bigl(\gamma_\beta^{-1}(s_*+)\bigr)\bigr)
=\rho_\beta^{-1}(s_*)
\end{gather*}
and 
\begin{gather*}
\beta\bigl(\gamma_\beta^{-1}(s^*)\bigr)=\rho_\beta^{-1}\bigl(\gamma_\beta\bigl(\gamma_\beta^{-1}(s^*)\bigr)\bigr)
=\rho_\beta^{-1}(s^*) .
\end{gather*}
In particular,
\begin{gather}\label{between}
\beta\bigl(\gamma_\beta^{-1}(s_*)+\bigr)=\rho_\beta^{-1}(s_*)\leq
\rho_\beta^{-1}(s)\leq
\rho_\beta^{-1}(s^*) =\beta\bigl(\gamma_\beta^{-1}(s^*)\bigr)
\end{gather}
and so
\begin{gather*}
\Gamma(\rho_\beta^{-1}(s^*),\gamma_\beta^{-1}(s)\bigr)=
\Gamma(\rho_\beta^{-1}(s_*),\gamma_\beta^{-1}(s)\bigr)\geq
\Gamma(z,\gamma_\beta^{-1}(s)\bigr) ,
\end{gather*}
by the definition of $\beta$.
With reference to Assumption \ref{Assumpmu} and Assumption \ref{Assumh}, it follows that $x\mapsto\Gamma(x;y)$
is concave. According to (\ref{between}), there exists $\lambda\in[0,1]$ such that
$\rho_\beta^{-1}(s)=\lambda\rho_\beta^{-1}(s_*)+(1-\lambda)\rho_\beta^{-1}(s^*)$.
Hence,
\begin{gather}%\label{Gcalc2}
\Gamma\bigl(\rho_\beta^{-1}(s);\gamma_\beta^{-1}(s)\bigr)
-\Gamma\bigl(z;\gamma_\beta^{-1}(s)\bigr)\geq
\Gamma\bigl(\beta\bigl(\gamma_\beta^{-1}(s)\bigr);\gamma_\beta^{-1}(s)\bigr)
-\Gamma\bigl(z;\gamma_\beta^{-1}(s)\bigr)\geq 0. 
\end{gather}
\end{proof}

\begin{proof}[\textbf{Proof of Lemma \ref{Lembetaprop}}]
%Next, we want to prove the claimed limiting behaviour of $\beta^*(y)$ as $y\rightarrow\bar{y}_A$. 
If $y>\bar{y}_A$, then by the definition of $\beta^*$, it holds that if 
$\lim_{x\rightarrow y-}\kappa_A(x)=\infty$ or $\lim_{x\rightarrow y-}\kappa_A'(x)=\infty$, 
then $\lim_{x\rightarrow y-}\beta^*(x)=\bar{z}$. The remaining case is when $y=\bar{y}_A$. We will 
prove this case by contradiction.  Suppose $\beta^*(\bar{y}_A)>\bar{z}$.
For any $x\in\bigl(\bar{z},\beta^*(\bar{y}_A)\bigr)$ 
and $y\in(0,\bar{y}_A)$ such that $\beta^*(y)\geq\beta^*(\bar{y}_A)$, we have
\begin{align*}
A\psi(x)&\leq A\biggl(\psi(x)-\psi\bigl(\beta^*(y)\bigr)\biggr)\cr  
&\leq\kappa_A(y)\biggl(\frac{1}{h\bigl(\beta^*(y)\bigr)}
-\frac{1}{h(x)}\biggr)+\kappa'_A(y)\biggl(H\bigl(\beta^*(y)\bigr)-H(x)\biggr)\cr%since0<y<\bar{y}_A,\beta^*(y) is maximiser
&\leq\kappa_A(y)\biggl(\frac{1}{h\bigl(\beta^*(\bar{y}_A)\bigr)}
-\frac{1}{h(x)}\biggr)+\kappa'_A(y)\biggl(H\bigl(\beta^*(\bar{y}_A)\bigr)-H(x)\biggr).%since\beta^*(y)\geq\beta^*(\bar{y}_A)
\end{align*}
Taking $y$ to be arbitrarily close to $\bar{y}_A$ implies $\psi(x)=-\infty$. 
This means $x<\bar{z}$, which contradicts $x>\bar{z}$. 
Hence, we conclude that $\beta^*(\bar{y}_A)=-\infty$.

Next we prove (\ref{lim-y-to-0}). Observe that if $\beta^*(0+)<0$, 
then (\ref{lim-y-to-0}) is true. However, if $\beta^*(0+)=0$, then 
\begin{gather*}
\frac{\kappa_A(y)}{h\bigl(\beta^*(y)\bigr)}
\,\geq\,\Gamma(x;y)-A\psi\bigl(\beta^*(y)\bigr)-\kappa'_A(y)H\bigl(\beta^*(y)\bigr)
\,\geq\,\Gamma(x;y)-\kappa'_A(y)H\bigl(\beta^*(y)\bigr),
\end{gather*}
from which it follows that for any $x\in(\bar{z},0)$, 
\begin{align}
0&\geq\liminf_{y\rightarrow 0+}\frac{\kappa_A(y)}{h\bigl(\beta^*(y)\bigr)}
\,\geq\, A\psi(x)-\limsup_{y\rightarrow 0+}\kappa'_A(y)H\bigl(\beta^*(y)\bigr),\label{liminf3}\\
0&\geq\limsup_{y\rightarrow 0+}\frac{\kappa_A(y)}{h\bigl(\beta^*(y)\bigr)}
\,\geq\, A\psi(x)-\liminf_{y\rightarrow 0+}\kappa'_A(y)H\bigl(\beta^*(y)\bigr).\label{limsup}
\end{align} 
Therefore, 
\begin{align*}
0&\geq\limsup_{y\rightarrow 0+}\kappa'_A(y)H\bigl(\beta^*(y)\bigr)
\,\geq\, A\psi(x),\cr
0&\geq\liminf_{y\rightarrow 0+}\kappa'_A(y)H\bigl(\beta^*(y)\bigr)
\,\geq\, A\psi(x).
\end{align*}
By letting $x$ tend to $0$, then with reference to (\ref{phipsicdt1}), we get 
$\lim_{y\rightarrow 0+}\kappa'_A(y)H\bigl(\beta^*(y)\bigr)=0$. 
Also, by letting $x$ tend to $0$ in (\ref{liminf3}) and (\ref{limsup}), 
$\lim_{y\rightarrow 0+}\frac{\kappa_A(y)}{h(\beta^*(y))}=0$ follows. 
\end{proof}

%---------------------------------------------------------------------

\begin{proof}[\textbf{Proof of Proposition \ref{Propv}}]
%\textbf{Continuity of $v$:}
%We want to prove $v$ is continuous, and first we show it is 
%finite. But with reference to (\ref{F_T-finite}), (\ref{Jdef}), 
%(\ref{JI(ii)(b)(c)}) and (\ref{JI(iii)}) it suffices to 
%show that with $\beta^*$ being given by Lemma \ref{LemGamma} 
%the strategy described in Lemma \ref{Lemmstrat} is admissible. 
To show that $v$ is continuous, we first prove it is finite. 
With reference to (\ref{tbar-infty=0})-(\ref{JI(iii)}), it is sufficent to 
show that the function $J_\beta$ given by (\ref{Jdef}) is finite for $\beta$ defined by Lemma \ref{LemGamma}. 
By the continuity of $Y^{\beta}$ and $Z^{Y^\beta}$ after time 0 and condition (\ref{phipsicdt4}), 
we have that there exists some $s>0$ such that 
\begin{gather} \label{Jbeta_finite_1}
\int_0^s \biggl(\kappa_A\bigl(Y_t^\beta\bigr)
+Ah\bigl(Z_t^{Y^\beta}\bigr)\psi\bigl(Z_t^{Y^\beta}\bigr)\biggr)\, dt<\infty 
\end{gather}
and $Y_s^\beta<\bar{y}_A$.
%Let $\beta=\beta^*(y)$ be the largest solution to (\ref{hequation}) 
%and let $Y^\beta$ denote the strategy corresponding to $\beta$ 
%as described in Lemma 
%\ref{Lemmstrat}. For any $y>0$ and $\epsilon>0$, 
%\begin{gather*}
%t_\epsilon=\inf\bigl\{t\geq 0\mid Y_t^{\beta}\leq\epsilon\bigr\}<\infty .
%\end{gather*}
%From Lemma \ref{LemGamma}, it follows that 
%$\kappa_A(Y_t^\beta)<\infty$, for every $t\geq 0$ 
%(if $\lim_{y\rightarrow\bar{y}_A}\kappa_A(y)=\infty$, then 
%$Y_t^\beta<\bar{y}_A$ for $t\geq 0$, and if 
%$\kappa_A(\bar{y}_A)<\infty$, then $Y^\beta_t\leq \bar{y}_A$ 
%for $t\geq 0$). Hence,
%\begin{gather*}
%\int_0^{t_\epsilon}\kappa_A\bigl(Y_t^\beta\bigr)\, dt<\infty .
%\end{gather*}
%Then in order to prove the admissibility of $Y^\beta$ it is sufficient to prove that
%\begin{gather*}
%\int_{t_\epsilon}^\infty \kappa_A\bigl(Y_t^\beta\bigr)\, dt<\infty .
%\end{gather*}
According to the condition in Lemma \ref{Lembetaprop}, 
\begin{gather*}
\lim_{y\rightarrow 0^+}\frac{\kappa_A(y)}{h\bigl(\beta(y)\bigr)}=0, 
\end{gather*}
so it follows that there exists $C_1>0$ and $0<\epsilon<\bar{y}_A$ such that
\begin{gather*}
\kappa_A(y)\leq -C_1 h\bigl(\beta(y)\bigr) ,\qquad\text{ for all } y\in[0,\epsilon]. 
\end{gather*}
Since $\psi\bigl(Z_t^{Y^\beta}\bigr)$ is bounded for all $t\geq s$ (it increases to 0), this and the above inequality 
imply that 
\begin{align} \label{Jbeta_finite_2}
\int_s^\infty \biggl(\kappa_A\bigl(Y_t^\beta\bigr)
+Ah\bigl(Z_t^{Y^\beta}\bigr)\psi\bigl(Z_t^{Y^\beta}\bigr)\biggr)\, dt
&\leq\int_s^\infty\bigg(-C_1 h\bigl(\beta(Y_t^\beta)\bigr)-C_2h\bigl(Z_t^{Y^\beta}\bigr)\bigg)\, dt\cr
&\leq\int_s^\infty\bigg(-C_1 h\bigl(Z_t^{Y^\beta}\bigr)-C_2h\bigl(Z_t^{Y^\beta}\bigr)\bigg)\, dt\cr
&\leq (C_1+C_2)\bigl(Y_s^\beta-Z_s^{Y^\beta}\bigr)<\infty, 
\end{align}
where $C_2>0$ is some constant. 
Therefore, (\ref{Jbeta_finite_1}) and (\ref{Jbeta_finite_2})  show that $v$ is finite. 

Note that both of the expressions (\ref{vdef1}) and (\ref{vdef2}) are 
continuous in $y$ and $z$. It is therefore sufficient to prove that $v$ 
is continuous across $\mathcal{G}^\beta$. 
Let $J_u(y,z)$ denote the expression for $v(y,z)$ given by (\ref{vdef1}), and let $J_l(y,z)$ 
denote the expression in (\ref{vdef2}). Suppose $(y,z)$ is a point on the graph of $\beta$, i.e, 
$z=\beta(y)$. Consider a sequence of points $(y_n,z_n)_{n=1}^\infty$ contained in 
$\overline{\mathcal{S}}^\beta\setminus\mathcal{G}^\beta$, converging to $(y,z)$. With 
reference to (\ref{JI(iii)calculation}) and(\ref{JI(iii)}), we calculate that
\begin{align}\label{cont-case}
\lim_{n\rightarrow \infty} v(y_n,z_n)
&=%\lim_{z\rightarrow \beta(y)^-}\biggl\{
%\kappa_A(y)\int_{\beta(y)}^z\frac{1}{h(u)}\, du+A\int_{z-y}^{\beta(y)-y}\psi(u)\, du
%-A\int_z^{\beta(y)}\psi(u)\, du\cr
%&\qquad+\int_0^{\beta(y)-y}\biggl(\frac{\kappa_A\bigl(\gamma_\beta^{-1}(u)
%\bigr)}{h\bigl(\rho_\beta^{-1}(u)\bigr)}+A\psi\bigl(\rho_\beta^{-1}(u)\bigr)
%-A\psi(u)\biggr)\, du\biggr\}\cr
J_u\bigl(y,\beta(y)\bigr)
=J_l\bigl(y,\beta(y)\bigr)
=v\bigl(y,\beta(y)\bigr) .
\end{align}
If $(y,z)$ lies on the graph of $\beta^{-1}$, i.e, $y=\beta^{-1}(z)$, 
then using the property that $\beta^{-1}(u)=\beta^{-1}(z)$, for $u\in\bigl(z,\beta(\beta^{-1}(z))\bigr)$, 
direct calculation results in (\ref{cont-case}). 
We therefore conclude that $v$ is a continuous function.
Differentiating $v$ gives  
\begin{align}
D_y^-v(y,z)&=-\frac{\kappa_A\bigl(\gamma_\beta^{-1}(z-y)
\bigr)}{h\bigl(\rho_\beta^{-1}(z-y)\bigr)}
-A\psi\bigl(\rho_\beta^{-1}(z-y)\bigr) ,
\quad\, &z>\beta(y),\label{D1}\\
v_z(y,z)&=\frac{\kappa_A\bigl(\gamma_\beta^{-1}(z-y)
\bigr)}{h\bigl(\rho_\beta^{-1}(z-y)\bigr)}
+A\psi\bigl(\rho_\beta^{-1}(z-y)\bigr) ,
\quad\, &z>\beta(y),\label{D2}\\
D_y^-v(y,z)&=\kappa_A'(y)H(z)
-\frac{\kappa_A(y)}{h\bigl(\beta(y)\bigr)}-A\psi\bigl(\beta(y)\bigr)
-\kappa_A'(y)H\bigl(\beta(y)\bigr) ,   &z\leq\beta(y) ,\label{D3}\\
v_z(y,z)&=\frac{\kappa_A(y)}{h(z)}+A\psi(z) ,\qquad
&z\leq\beta(y) .\label{D4}
\end{align}
These expressions are left-continuous with right limit in $y$ and continuous in $z$ 
(all of these expressions are continuous at $(0,0)$, which is guaranteed by (\ref{lim-y-to-0})). 
%((although they could be infinite at some end points)) 
Also, we check that 
for any $(y_n,z_n)_{n=1}^{\infty}\subseteq\overline{\mathcal{S}}^{\beta}$, $(y,z)\in\mathcal{G}^{\beta}$ and 
$\lim_{n\rightarrow\infty}(y_n,z_n)=(y,z)$, we have $v_z(y_n,z_n)\rightarrow v_z(y,z)$, as $n\rightarrow\infty$. 
Further,  $\lim_{z\rightarrow\beta(y)+}D_y^-v(y,z)=D_y^-v\bigl(y,\beta(y)\bigr)$. 
Therefore, we conclude that $v_z(y,z)$ is continuous, and $D_y^-v(y,z)$ is c\`{a}gl\`{a}d in $y$ and continuous in $z$. 

Standard calculations show that $v$ satisfies 
(\ref{h1}) and (\ref{h3}). When $z=0$, (\ref{h2}) is clearly true. 
In order to verify (\ref{h2}) for $z\neq 0$, we compute that when $z>\beta(y)$, 
\begin{align}\label{temp}
&h(z)v_z(y,z)-\kappa_A(y)-A h(z)\psi(z)\cr
=\,&h(z)\biggl\{\frac{\kappa_A\bigl(\gamma_\beta^{-1}(s)
\bigr)}{h\bigl(\rho_\beta^{-1}(s)\bigr)}-\frac{\kappa_A(z-s)}{h(z)}
+A\bigl\{\psi\bigl(\rho_\beta^{-1}(s)\bigr)-\psi(z)\bigr\}\biggr\}, 
\end{align}
where $s=z-y$. Observe that $h\bigl(\rho_\beta^{-1}(s)\bigr)=0$ implies $y=0$, 
but (\ref{h1})--(\ref{h4}) are under the condition that $y>0$. So $h\bigl(\rho_\beta^{-1}(s)\bigr)$ is non-zero. 
By the definition 
of $\gamma_\beta^{-1}$, we must have $\gamma_\beta^{-1}(s)\in(0,\bar{y}_A)$ if 
$\beta(\bar{y}_A)=\bar{z}$, or $\gamma_\beta^{-1}(s)\in(0,\bar{y}_A]$ if 
$\beta(\bar{y}_A)>\bar{z}$. Then according to 
the limiting behaviour of $\beta$ in Lemma \ref{Lembetaprop}, $\kappa_A\bigl(\gamma_\beta^{-1}(s)\bigr)$ 
must be finite, and therefore also  $\kappa'_A\bigl(\gamma_\beta^{-1}(s)\bigr)$. However, $\kappa_A(z-s)$ 
may be infinite, but then it follows that (\ref{temp}) is negative. Otherwise, if 
$\kappa_A(y)<\infty$, write
\begin{gather*}
G(s;z)=\frac{\kappa_A\bigl(\gamma_\beta^{-1}(s)
\bigr)}{h\bigl(\rho_\beta^{-1}(s)\bigr)}-\frac{\kappa_A(z-s)}{h(z)}
+A\bigl\{\psi\bigl(\rho_\beta^{-1}(s)\bigr)-\psi(z)\bigr\}. 
\end{gather*}
Then in order to verify (\ref{h2}), it is sufficient to show that $G(s;z)\geq 0$, for all 
$\rho_\beta^{-1}(s)<z<0$. We calculate that $G(s;y)$ can be expressed by 
\begin{align} \label{Gcalc1}
G(s;z)&=\Bigl[\Gamma\bigl(\rho_\beta^{-1}(s);\gamma_\beta^{-1}(s)\bigr)
-\Gamma\bigl(z;\gamma_\beta^{-1}(s)\bigr)\Bigr]\cr
&\,-\kappa'_A\bigl(\gamma_\beta^{-1}(s)\bigr)\Bigl[H\bigl(\rho_\beta^{-1}(s)\bigr)\bigr)-H(z)\Bigr]
+\frac{1}{h(z)}\biggl[\kappa_A\bigl(\gamma_\beta^{-1}(s)\bigr)-\kappa_A(z-s)\biggr], 
\end{align}
%where Lemma \ref{LemGamma} verifies 
According to Lemma \ref{Gammaconvex},
\begin{gather*}
\Gamma\bigl(\rho_\beta^{-1}(s);\gamma_\beta^{-1}(s)\bigr)
-\Gamma\bigl(z;\gamma_\beta^{-1}(s)\bigr)\geq 0 .
\end{gather*}
Furthermore, we calculate that
\begin{align}\label{Gcalc3}
&\frac{1}{h(z)}\Bigl[\kappa_A\bigl(\gamma_\beta^{-1}(s)\bigr)-\kappa_A(z-s)\Bigr]
-\kappa'_A\bigl(\gamma_\beta^{-1}(s)\bigr)\Bigl[H\bigl(\rho_\beta^{-1}(s)\bigr)-H(z)\Bigr]\cr
=\,&\int_{\rho_\beta^{-1}(s)}^{z} 
\Bigg(\frac{\bigl[\kappa_A(u-s)-\kappa_A\bigl(\rho_\beta^{-1}(s)-s\bigr)\bigr]h'(u)}{h^2(u)}
+\frac{\kappa'_A\bigl(\rho_\beta^{-1}(s)-s\bigr)-\kappa'_A(u-s)}{h(u)}\Bigg) \,du\cr
\geq\,&0.
\end{align}
(\ref{h2}) then follows from (\ref{Gcalc1})--(\ref{Gcalc3}). Moreover, from the definition 
of $\beta$ we get 
\begin{align*} %\label{uyuz}
D_y^-v(y,z)+v_z(y,z)&=\kappa_A'(y)H(z)-\frac{\kappa_A(y)}{h\bigl(\beta(y)\bigr)}
+A\psi(z-y)-A\psi\bigl(\beta(y)\bigr)\cr
&\qquad-\kappa_A'(y)H\bigl(\beta(y)\bigr)
+\frac{\kappa_A(y)}{h(z)}+A\psi(z)-A\psi(z-y)\cr
&=\Gamma(z;y)-\Gamma\bigl(\beta(y);y\bigr)\cr
&\leq 0 ,
\end{align*}
which verifies (\ref{h4}). 

Finally, the expression in (\ref{vdef1}) satisfies the boundary condition since, 
for any $u\in[\beta(0+),z]$, we have $\gamma_\beta^{-1}(u)=0$ and $\rho_\beta^{-1}(u)=u$. 
The expression in  (\ref{vdef2}) clearly satisfies the boundary condition. 
\end{proof}

%------------------------------------------------------------------------

\begin{proof}[\textbf{Proof of Theorem \ref{Thmmain}}]
Let $\delta$ be a positive-valued $C^{\infty}(\mathbb{R})$ function
with support on $[0,1]$ satisfying $\int_0^1 \delta(x)\, dx=1$,
and define a sequence of functions $\{\delta_n\}_{n=1}^\infty$
by
\begin{gather*}
\delta_n(s)=n\, \delta(n s) , \quad s\geq 0 .
\end{gather*}
We mollify $v$ to obtain a
sequence of functions $\{v^{(n)}\}_{n=1}^\infty$ which are given by
\begin{gather*}
v^{(n)}(y,z)=\int_0^1 v(y-s,z)\, \delta_n(s)\, ds. 
\end{gather*}
(One may extend the lower bound of the domain of $v(\cdot,z)$ properly so that $v^{(n)}$ is well-defined at $y=0$.)
Then $v^{(n)}\in C^{1,1}(\mathcal{D})$, for all $n\in\mathbb{N}$, and
\begin{align*}
v(y,z)&=\lim_{n\rightarrow\infty}v^{(n)}(y,z) ,\cr
v_z(y,z)&=\lim_{n\rightarrow\infty} v_z^{(n)}(y,z),\cr
D_y^-v(y,z)&=\lim_{n\rightarrow\infty} v_y^{(n)}(y,z),
\end{align*}
where the last equality is due to $D_y^-v(y,z)$ being \gd in $y$. Moreover, for 
every $(y_0,z_0)\in\mathcal{D}$ there exists a $K>0$ such that 
on the set $\bigl\{\,(y,z)\in\mathcal{D}\,\big|\,z\geq y+z_0-y_0\,\bigr\}$, 
\begin{align}\label{sup1}
&\bigl\arrowvert
v^{(n)}(y,z)\bigr\arrowvert\leq K , \quad  n\in\mathbb{N} ,\\ \label{sup2}
&\bigl\arrowvert v_y^{(n)}(y,z)\bigr\arrowvert\leq K ,
\quad   n\in\mathbb{N} ,\\
&\bigl\arrowvert
v_z^{(n)}(y,z)\bigr\arrowvert\leq K ,
\quad   n\in\mathbb{N} .\label{sup3}
\end{align}
(If $Y$ is admissible and $(Y_{0-},Z^Y_{0-})=(y_0,z_0)$, 
then $(Y_t,Z^Y_t)\in\bigl\{\,(y,z)\in\mathcal{D}\,\big|\,z\geq y+z_0-y_0\,\bigr\}$, for all $t\geq 0$.)
By It\^{o}'s formula, we calculate that 
\begin{align}  \label{itovn}
&v^{(n)}\bigl(Y_T,Z_T^Y\bigr)+\int_0^T\biggl(\kappa_A\bigl(Y_{t-}\bigr)+Ah\bigl(Z_{t-}^Y\bigr)
\psi\bigl(Z_{t-}^Y\bigr)\biggr)\, dt\cr
\quad&=v^{(n)}(y,z)
+\int_0^T\biggl(v_y^{(n)}(Y_{t-},Z_{t-}^Y)+v_z^{(n)}(Y_{t-},Z_{t-}^Y)\biggr)\, dY_t^c\cr
&\qquad+\int_0^T\biggl(\kappa_A\bigl(Y_{t-}\bigr)+Ah\bigl(Z_{t-}^Y\bigr)\psi\bigl(Z_{t-}^Y\bigr)
-v_z^{(n)}\bigl(Y_{t-},Z_{t-}^Y\bigr)h\bigl(Z_{t-}^Y\bigr)\biggr)\, dt\cr
&\qquad+\sum_{0\leq t\leq T}\biggl\{v^{(n)}\bigl(Y_{t-}+\triangle Y_t, Z_{t-}^Y+\triangle Y_t\bigr)
-v^{(n)}\bigl(Y_{t-},Z_{t-}^Y\bigr)\biggr\},
\end{align}
for all $Y\in\mathcal{A}_D(y)$. Observe that for $t\geq 0$,
\begin{gather*}
0\leq-\int_0^th\bigl(Z_{u}^Y\bigr)\,du=Z_{t}^Y-Y_t-Z_0^Y+Y_0\leq y-z.
\end{gather*}
Then, with reference to (\ref{sup1})--(\ref{sup3}), we have 
\begin{align*}
\int_0^\infty \sup_{n\in\mathbb{N}}\Bigl\arrowvert v_z^{(n)}
\bigl(Y_{t-},Z_{t-}^Y\bigr)h\bigl(Z_{t-}^Y\bigr)\Bigr\arrowvert\, dt\leq K(y-z). 
\end{align*}
Similarly,
\begin{align*}
\int_0^\infty \sup_{n\in\mathbb{N}}\Bigl\arrowvert
v_y^{(n)}\bigl(Y_{t-},Z_{t-}^Y\bigr)
+v_z^{(n)}\bigl(Y_{t-},Z_{t-}^Y\bigr)\Bigr\arrowvert
\, d(-Y_{t}^c)\leq 2Ky
\end{align*}
and
\begin{align*}
\sum_{0\leq t}\sup_{n\in\mathbb{N}} \Bigl\arrowvert
v^{(n)}\bigl(Y_{t-}+\triangle Y_t,Z_{t-}^Y
+\triangle Y_t\bigr)-v^{(n)}\bigl(Y_{t-},Z_{t-}^Y\bigr)\Bigr\arrowvert
\leq 2K y .
\end{align*}
Hence, by (\ref{itovn}) and the boundary condition $v(0,z)=A\int_0^z\psi(u)\,du$, 
it follows from
the dominated convergence theorem that for any $Y\in\mathcal{A}_D(y)$,
\begin{align}\label{itov}
&\int_0^\infty\biggl(\kappa_A\bigl(Y_{t-}\bigr)+Ah\bigl(Z_{t-}^Y\bigr)
\psi\bigl(Z_{t-}^Y\bigr)\biggr)\, dt\cr
=\,&v(y,z)+\int_0^\infty\biggl(D_y^-v\bigl(Y_{t-},Z_{t-}^Y\bigr)
+v_z\bigl(Y_{t-},Z_{t-}^Y\bigr)\biggr)\, dY_t^c\cr
&\quad+\int_0^\infty\biggl(\kappa_A\bigl(Y_{t-}\bigr)+Ah\bigl(Z_{t-}^Y\bigr)\psi\bigl(Z_{t-}^Y\bigr)
-v_z\bigl(Y_{t-},Z_{t-}^Y\bigr)h\bigl(Z_{t-}^Y\bigr)\biggr)\, dt\cr
&\quad+\sum_{t\geq 0}\biggl\{v\bigl(Y_{t-}+\triangle Y_t, Z_{t-}^Y+\triangle Y_t\bigr)
-v\bigl(Y_{t-},Z_{t-}^Y\bigr)\biggr\}, 
\end{align}
as $n\rightarrow\infty$ and $T\rightarrow\infty$. According to Proposition 
\ref{Propv}, $v$ satisfies (\ref{h1})--(\ref{h4}), and therefore,
\begin{gather}\label{vineq3}
\int_0^\infty\biggl(\kappa_A\bigl(Y_{t-}\bigr)+Ah\bigl(Z_{t-}^Y\bigr)
\psi\bigl(Z_{t-}^Y\bigr)\biggr)\, dt
\geq v(y,z) .
\end{gather}
Hence, $V\geq v$.

From from (\ref{Jbeta_finite_1})-(\ref{Jbeta_finite_2}), 
we know that with $\beta$ being the largest solution to (\ref{hequation}) 
and $Y^{\beta}$ being the strategy described in Lemma \ref{Lemmstrat} 
corresponding to $\beta$, $Y^{\beta}$ is admissible, in particular (\ref{stratAssump1}) is satisfied. Therefore, 
with reference to (\ref{vineq3}), in order to complete the 
proof, we need to show that (\ref{vineq3}) holds with equality 
for $Y^{\beta}$. Observe that $\triangle Y^{\beta}<0$ only if 
$t=0$ and $z>\beta(y)$. But by (\ref{h1}) and Proposition \ref{Propv}, 
we have that $D_y^-v(y,z)+v_z(y,z)=0$, 
for $z>\beta(y)$. Therefore,
\begin{gather*} %\label{jump}
\sum_{t\geq 0} \biggl\{v\bigl(Y_{t-}^{\beta}+\triangle Y_t^{\beta},Z_{t-}^{Y^{\beta}}
+\triangle Y_t^{\beta}\bigr)
-v\bigl(Y_{t-}^{\beta},Z_{t-}^{Y^{\beta}}\bigr)\biggr\}=0 .
\end{gather*}
For any $z\leq 0$, if $0\leq t\leq t_w$,
where $t_w$ is defined by (\ref{twdef}), then $d\bigl(Y_t^\beta\bigr)^c=0$, hence 
\begin{gather*}
\int_0^{t_w}\biggl(D_y^-v\bigl(Y^\beta_{t-},Z_{t-}^{Y\beta}\bigr)
+v_z\bigl(Y^\beta_{t-},Z_{t-}^{Y^\beta}\bigr)\biggr)\, d\bigl(Y^\beta_t\bigr)^c=0; 
\end{gather*}
if $t>t_w$, then $\bigl(Y_{t}^{\beta},Z_{t}^{Y^{\beta}}\bigr)\in\mathcal{G}^\beta$, which implies 
\begin{align*}
\int_{t_w}^\infty\biggl(D_y^-v\bigl(Y_{t-}^{\beta},Z_{t-}^{Y^{\beta}}\bigr)
+v_z\bigl(Y_{t-}^{\beta},Z_{t-}^{Y^{\beta}}\bigr)\biggr)
\, d(Y_{t}^{\beta})^c=0. 
\end{align*}
Finally we have
\begin{align*}
\int_{0}^\infty\biggl(\kappa_A\bigl(Y_{t-}^{\beta}\bigr)+Ah\bigl(Z_{t-}^{Y^{\beta}}\bigr)
\psi\bigl(Z_{t-}^{Y^{\beta}}\bigr)
-v_z\bigl(Y_{t-}^{\beta},Z_{t-}^{Y^{\beta}}\bigr)h\bigl(Z_{t-}^{Y^{\beta}}\bigr)\biggr)\, dt=0, 
\end{align*}
since the integrand is equal to $0$, for all 
$\bigl(Y_{t}^{\beta},Z_{t}^{Y^{\beta}}\bigr)\in\overline{\mathcal{W}}^\beta$, 
and the Lebesgue measure of the set of $t\geq 0$ for which  
$\bigl(Y_{t}^{\beta},Z_{t}^{Y^{\beta}}\bigr)\in\overline{\mathcal{S}}^\beta_s\setminus\mathcal{G}^\beta$ 
is $0$. With reference to (\ref{itov}), we therefore conclude that $v=V$ and that 
$Y^*=Y^{\beta}\in\mathcal{A}_D(y)$ is an admissible optimal liquidation strategy 
for the optimization problem (\ref{Vdef2}), and the result follows from (\ref{red-eq}). 
\end{proof}

%==================================================================

\bibliographystyle{apalike} 
\bibliography{references}

\end{document}